\documentclass[preprint,12pt,authoryear]{elsarticle}

\usepackage{amsmath,amssymb,graphicx}
\usepackage{xcolor}
\usepackage[defaultcolor=red]{changes}
\usepackage{booktabs,multirow}
\usepackage[hidelinks]{hyperref}
\usepackage{siunitx}
\usepackage[utf8]{inputenc}
\usepackage{newunicodechar}
\newunicodechar{∼}{\textasciitilde}
\usepackage{pdflscape}
\usepackage{longtable,booktabs,multirow}

\journal{Journal of High Energy Astrophysics}
\newcommand{\src}{1A~1246$-$588}

\newcommand{\astr}{\emph{AstroSat}}
\newcommand{\nicer}{\emph{NICER}}

\begin{document}

\begin{frontmatter}

\title{A \nicer\ and \astr\ view of the neutron star low-mass X-ray binary \src\ }

\author[inst1,inst2]{Vaidehi Poojyam\corref{cor1}\fnref{orcid1}}
\ead{<vaidehi.poojyam15@gmail.com>}
\author[inst2]{{Vikas Mistry}\fnref{orcid2}}
\author[inst3,inst4]{Yash Bhargava\fnref{orcid3}}
\author[inst3]{Sudip Bhattacharyya\fnref{orcid4}}
\author[inst2]{Nitinkumar Bijewar}

\cortext[cor1]{Corresponding author.}
\fntext[orcid1]{ORCID: \href{https://orcid.org/0009-0006-5654-1473}{0009-0006-5654-1473}}
\fntext[orcid2]{ORCID: \href{https://orcid.org/0009-0009-6459-0728}{0009-0009-6459-0728}}
\fntext[orcid3]{ORCID: \href{https://orcid.org/0000-0002-5967-8399}{0000-0002-5967-8399}}
\fntext[orcid4]{ORCID: \href{https://orcid.org/0000-0002-6351-5808}{0000-0002-6351-5808}}

\affiliation[inst1]{organization={Department of Physics and Astronomy, University of Alabama}, city={Tuscaloosa}, state={AL},
            % postcode={35404}, 
            country={USA}}
            
\affiliation[inst2]{organization={Department of Physics, University of Mumbai}, city={Mumbai}, 
% postcode={400098}, 
country={India}}

\affiliation[inst3]{organization={Department of Astronomy and Astrophysics, Tata Institute of Fundamental Research}, 
% addressline={1 Homi Bhabha Road, Colaba}, 
city={Mumbai}, 
% postcode={400005}, 
country={India}}

\affiliation[inst4]{organization={INAF–Osservatorio Astronomico di Cagliari}, 
            % addressline={Via della Scienza, 5},
            city={Selargius},
            % postcode={09047},
            country={Italy}
}

\begin{abstract}

Neutron star (NS) low-mass X-ray binary (LMXB) systems depict a variety of X-ray spectral and timing features, which can be useful to probe the accretion-ejection mechanism in the strong gravity regime.
Here, we study the relatively unexplored and faint NS LMXB \src, which is also an ultra-compact X-ray binary (UCXB) with a white dwarf donor. We investigate its temporal and spectral behavior using pointed \nicer\ and \astr\ observations, supported by long-term \textit{MAXI}/GSC monitoring.
The \textit{MAXI} light curve shows modest, recurrent outburst-like enhancements, providing the long-term flux context for interpreting the pointed observations. During the \astr\ observations in 2017, the source exhibits an absorbed 0.4–20~keV flux of $(1.18\pm0.02)\times10^{-10}$~erg~cm$^{-2}$~s$^{-1}$, while during the \textit{NICER} observations in 2019 it spans an absorbed 0.5–10~keV flux range of $(0.7$–$3.7)\times10^{-10}$~erg~cm$^{-2}$~s$^{-1}$ and traces an atoll-like pattern in the hardness–intensity diagram.
Broadband spectral modeling shows that the emission is well described by a soft blackbody and a hard Comptonized component, with no statistically required multicolor disk contribution. The blackbody temperature increases from  0.28 to $0.39$~keV, with an emitting radius consistent within 6.9--$13.8$~km, while the Comptonization photon index varies from $1.8$ to $2.3$. We find that the observed spectral-state evolution is driven by a redistribution of accretion power between thermal emission from the NS boundary layer and Comptonized emission, consistent with atoll-type behavior. These results provide the first quantitative, multi-epoch view of accretion-state evolution in \src, revealing systematic changes in the thermal boundary-layer emission and the Comptonizing region in this UCXB system.
\end{abstract}

\end{frontmatter}

% \linenumbers  % enable if using line numbers
% ...

\bibliographystyle{elsarticle-harv}


% \keywords{}


\section{Introduction} \label{sec:intro}

Low-mass X-ray binaries (LMXBs) are stellar systems with a low-mass ($\lesssim 1M_\odot$) donor/companion star and a compact object, the latter being either a black hole (BH) or a neutron star (NS). In an LMXB, the donor star fills its Roche lobe and its matter flows through the inner Lagrangian point to the compact object via an accretion disk 
\citep{bhattacharya1991PhR...203....1B}.
The donor can fill its Roche lobe when its radius increases due to evolution or irradiation, or if the Roche lobe size decreases when the binary orbit shrinks. 
The latter case is common for ultra-compact X-ray binaries (UCXBs), which have a typical period of $\lesssim 80$~min and have a dwarf donor star. 
In most cases, a UCXB hosts an NS as the compact object \citep[see][for a complete list of known UCXBs to date]{ArmasPadilla2023A&A}. In NS LMXBs, the classification of spectral states is traditionally based on the source trajectories in hardness–intensity and color–color diagrams and the relative dominance of spectral components.
Based on the shape traced on their hardness-intensity diagrams (HID) or color-color diagrams (CCD), the NS LMXBs are roughly classified into two types: Z sources and atoll sources \citep{1989A&A...225...79H}. Z sources accrete at a higher accretion rate with an X-ray luminosity of $\gtrsim 0.5~L_{Edd}$ and trace a Z-shaped track in their HID or CCD. Atoll sources span a broader range of luminosities, typically between $10^{-3}~L_{Edd}$ and $0.5~L_{Edd}$ \citep{2004astro.ph.10551V, Done_2007}.  % 
Most of the atoll sources are characterized by two distinct tracks in their HIDs and CCDs,
a banana-like structure and isolated patches called island states \citep{1989A&A...225...79H, 2004astro.ph.10551V, 2006A&A...460..233C}. 
Furthermore, the banana branch is often divided into upper and lower banana based on their hardness and intensity, and these two states exhibit different spectral and timing properties \citep{1989A&A...225...79H,2004astro.ph.10551V}. Within this broader phenomenology, the standard LMXB spectral frameworks offer two widely used interpretations. In the {\it Eastern} model \citep{mitsuda1984, mitsuda1989}, the soft X-ray emission arises from a multi-temperature accretion disk, while the hard component is produced by Comptonization in the inner disk region. In contrast, the {\it Western} model \citep{ mitsuda1984, white1988} attributes the soft emission to blackbody radiation from the neutron star surface or boundary layer, with the hard component generated in an optically thick Comptonizing region associated with the boundary layer.

\src\ is a relatively poorly studied LMXB, which was discovered as a persistent X-ray source in the Ariel survey \citep{carpenter1977MNRAS.179P..27C}.
The source underwent thermonuclear X-ray bursts that last about $\sim$150~s \citep{piro1997IAUC.6538....2P, boller1997IAUC.6546....1B, Zand2008A&A...485..183I}, confirming that the compact object is an NS.
\citet{Bassa2006A&A...446L..17B} identified the companion star in the system as a white dwarf (WD), strongly suggesting that \src\ is a UCXB \citep{asai2022PASJ...74..974A}, although the orbital period is still unknown \citep{Zand2008A&A...485..183I}. The distance of the source based on the optical properties has been estimated to be about 4.3~kpc \citep{Zand2008A&A...485..183I}. The source has also shown kilohertz (kHz) quasi-periodic oscillations (QPO) at $\sim$1260~Hz \citep{jonker2007MNRAS.378.1187J, Doesburgh2018MNRAS.479..426V}. \citet{Zand2008A&A...485..183I} investigated the nature of bursts and the persistent emission using observations from \textit{BeppoSAX}, \textit{RXTE}/PCA, \textit{Swift}/XRT, and \textit{XMM-Newton}. The authors 
reported that the regular PCA monitoring of the source demonstrates low and high flux states that differ in spectral composition. 
The 3--20~keV spectra from \textit{RXTE}/PCA of the low state were modeled sufficiently with a power-law-like emission, and the high-state emission in 0.3--20~keV (\textit{Swift}/XRT and \textit{RXTE}/PCA) showed a significant high-energy cutoff and was modeled as an absorbed Comptonization ($kT_{\rm e}\approx2.4$~keV) with an absorption column density of $2.1\times10^{21}$~cm$^{-2}$. 
The long-term ASM light curve of \src\ binned at 7~days showed aperiodic variations at time scales of $\sim100$~days \citep{Zand2008A&A...485..183I}, and the weekly PCA monitoring hinted that these long-term variations relate to the high and low states of the source. \citet{asai2022PASJ...74..974A} modeled the long-term variation of the source and interpreted the quasi-periodic variation at time scales of years as various slow mass-transfer processes from donor to the compact object (e.g., variation in the mass-transfer rate, variation in the irradiation from the compact object to the donor, etc.). 

A key question that remains unexplored for \src\ is how its soft X-ray spectral properties evolve on day timescales, something that earlier broadband studies could not resolve. Neutron Star Interior Composition Explorer's (\nicer) sensitivity in the 0.5--10~keV band allows us to track these short-term spectral changes and connect them to the source’s accretion state evolution. \astr~observations provide complementary broadband coverage with its simultaneous Soft X-ray Telescope (SXT; 0.3--8 keV) and Large Area X-ray Proportional Counter (LAXPC; 3--80 keV) observations, allowing us to probe the hard X-ray spectral shape and place constraints on the presence of high-energy cutoffs that are inaccessible to \nicer. Even when the hard X-ray statistics are limited, the \astr\ data help place the \nicer\ soft X-ray spectral evolution in the context of the source’s overall accretion state.
In this paper, we characterize the evolution of the spectral properties of \src\ using \nicer\ (2019) and \astr\ (2017) observations and interpret these changes in the context of the long-term, recurrent variability revealed by the Monitor of All-sky X-ray Image (\textit{MAXI}; \citet{MAXI2009PASJ...61..999M}).  
In Section~\ref{sec:obs}, we describe the observation and data reduction methods used to extract the spectral products. In Section~\ref{sec:analysis}, we detail the analysis methods, and in Section~\ref{sec:res} and \ref{sec:evol_spectral_&_accretion}, we discuss the results and interpret our findings. 

\section{Observation and Data reduction}\label{sec:obs}

\subsection{\nicer} \label{sec:nicer}

\nicer\ \citep{Gendreu2016} X-ray Timing Instrument \citep[XTI;][]{2014SPIE.9144E..20A} observed \src\ over a span of 72~days from 21 November, 2019 to 31 January, 2020. The corresponding observation IDs are listed in Table~\ref{tab:obs_log}. For our analysis, we use data in the 0.5--10~keV range, consistent with the well-calibrated energy band of the \nicer-XTI. Data reduction and extraction of spectral and timing products are performed using the \nicer\ CALDB version \texttt{xti20240206}. Data from detectors 14 and 34 were excluded from the analysis owing to their known elevated instrumental noise. Observations with a net exposure time shorter than or equal to 50~s (e.g., ObsID 2200970127; see Table \ref{tab:obs_log}) were excluded from further analysis, as such exposures do not provide sufficient statistics for reliable spectral or timing measurements. The raw level 1 data are processed into level 2 event files using the \texttt{nicerl2} pipeline. Spectra and associated Response Matrix File (RMF) and Ancillary Response File (ARF) are extracted using \texttt{nicerl3-spect}, while light curves across various energy bands are generated with \texttt{nicerl3-lc}. To model and subtract the background, we employ the SCORPEON model (NASA/GSFC 2023)\footnote{\url{https://heasarc.gsfc.nasa.gov/docs/nicer/analysis_threads/scorpeon-overview/}} which accounts for the non X-ray background and the cosmic X-ray background, utilizing the appropriate file-based background estimator.
Preliminary light curves reveal occurrences of apparent flaring activity. To determine whether these are instrumental in origin, such as precipitation electron flares, we examine the high-energy (8--12~keV) intensity, overshoot rate, and the geomagnetic cutoff rigidity parameter (\texttt{COR\_SAX})\footnote{\url{https://heasarc.gsfc.nasa.gov/docs/nicer/analysis_threads/flares/}}. Noting a correlation, we exclude intervals with \texttt{COR\_SAX} $<$ 1.5~GeV/$c$ and  eliminate the flaring features, thereby confirming their instrumental nature.
Following the removal of flaring intervals, all light curve and spectral products are re-extracted. The light curve for one of the observations (ObsID 2200970129) depicts a thermonuclear burst-like structure. As the primary  focus of the work is the persistent emission properties of \src\ in the soft X-rays, we exclude this observation from our analysis; however, such bursts have been reported for \src\ \citep{piro1997IAUC.6538....2P,  boller1997IAUC.6546....1B, Zand2008A&A...485..183I}. The final background-subtracted light curve is shown in Figure~\ref{fig:LC}. For normalization purposes, the intensities are scaled relative to the Crab, using data from a \nicer\ observation conducted on 26 November, 2019 (ObsID 2205010101).

\subsection{\textit{AstroSat}} \label{sec:astr}
\astr\ \citep{Singh2014SPIE.9144E..1SS} observed \src\ from 8 April, 2017 to 9 April, 2017 
(observation ID G07\_065T02\_9000001150). 
The details of the observation are mentioned in the Table~\ref{tab:obs_log}. 
We use the data from two primary instruments aboard \astr: the Soft X-ray Telescope \citep[SXT;][]{singh2016SPIE.9905E..1ES, Singh2017JApA...38...29S} and the Large Area X-ray Proportional Counter  \citep[LAXPC;][]{Yadav2016SPIE.9905E..1DY, Yadav2017CSci..113..591Y}. We investigate the spectral properties of the source in 0.4--20~keV,\ where the SXT calibration is robust and the LAXPC signal is not dominated by background, using the data products extracted from the SXT and LAXPC instruments. In order to compare the source properties during the epochs of \astr\ and \nicer\ observations, we normalize the intensity using a nearby Crab observation (observation ID A02\_090T01\_9000000970 taken on 21 January, 2017). No thermonuclear bursts are detected during the \astr\ observation, indicating that the source remains in a persistent, non-bursting state throughout this epoch.

\begin{table}[htbp]
\centering
\caption{\nicer\ and \astr\ observation log for \src.}
\label{tab:obs_log}
\begin{tabular}{l|l|l|l|r}
\toprule
Telescope & Instrument & ObsID & {MJD start} & {Exposure (ks)} \\
\midrule

\multirow{28}{*}{\nicer} & \multirow{28}{*}{XTI}
& 2200970101 & 58808.139 & 4.81 \\
& & 2200970102 & 58809.170 & 4.22 \\
& & 2200970103 & 58810.008 & 9.16 \\
& & 2200970104 & 58811.046 & 5.11 \\
& & 2200970105 & 58812.072 & 2.10 \\
& & 2200970106 & 58813.171 & 3.72 \\
& & 2200970107 & 58814.008 & 4.60 \\
& & 2200970108 & 58815.245 & 3.38 \\
& & 2200970122 & 58833.946 & 0.87 \\
& & 2200970123 & 58834.010 & 2.59 \\
& & 2200970124 & 58838.064 & 2.88 \\
& & 2200970125 & 58843.045 & 0.46 \\
& & 2200970126 & 58847.363 & 0.77 \\
& & 2200970127 & 58849.536 & 0.05 \\
& & 2200970128 & 58850.502 & 2.72 \\
& & 2200970130 & 58851.994 & 4.14 \\
& & 2200970131 & 58853.207 & 2.95 \\
& & 2200970132 & 58854.434 & 1.92 \\
& & 2200970133 & 58855.400 & 1.77 \\
& & 2200970134 & 58856.369 & 2.49 \\
& & 2200970136 & 58858.046 & 1.38 \\
& & 2200970137 & 58859.077 & 1.75 \\
& & 2200970138 & 58860.174 & 2.37 \\
& & 2200970139 & 58861.270 & 1.57 \\
& & 2200970140 & 58862.883 & 3.53 \\
& & 2200970141 & 58863.270 & 3.52 \\
& & 2200970142 & 58872.187 & 1.35 \\
& & 2200970143 & 58879.650 & 0.59 \\
\midrule
\multirow{2}{*}{\astr} & SXT   & \multirow{2}{*}{G07\_065T02\_9000001150} & \multirow{2}{*}{57851.831} & 25.22 \\
                       & LAXPC &                                          &                            & 40.30 \\
\bottomrule
\end{tabular}
\end{table}

\subsubsection{SXT}
\label{sec:sxt} % used for referring to this section from elsewhere

SXT level~2 data are downloaded from the Indian Space Science Data Centre (ISSDC) data archive. The data from multiple orbits are merged using the Julia tool \texttt{sxtevtmergerjl}\footnote{\url{https://www.tifr.res.in/~astrosat_sxt/dataanalysis.html}}.
For the analysis, a region of $15^\prime$ centered on the source is selected using \texttt{ds9}, and the \texttt{xselect} tool is used to extract the light curve and spectrum of the source. The exposure time of SXT is 25.22~ks with a net intensity of 1.363~$\pm$~0.007~counts~s$^{-1}$. At these intensities, pile-up effects are negligible, and thus we use a circular region for extraction. The \texttt{sxtARFModule} tool is utilized to modify the standard ARF supplied by the SXT Payload Operation Centre (POC), and the spectrum is grouped according to the detector response using \texttt{ftgrouppha}. The standard response and background files provided by the SXT POC are used for the analysis. The SXT spectrum is restricted to 0.4--7.0~keV because of calibration uncertainties at lower energies and poor signal-to-noise at higher energies, where the effective area declines sharply.

\subsubsection{LAXPC}

The level~1 LAXPC data are processed and converted to level~2 data using the \texttt{laxpc\textunderscore make\textunderscore event} tool from the pipeline \texttt{LAXPCsoftware22Aug15}\footnote{\url{http://astrosat-ssc.iucaa.in/laxpcData}}, with the standard filtering applied to exclude the South Atlantic Anomaly and the Earth occultation intervals using the \texttt{laxpc\_make\_stdgti} tool. 
We extract spectra from LXP10 and LXP20 as the LXP30 unit had a gas leak early in the mission.
The LAXPC observation is background dominated, and thus we extract the spectrum and the background spectrum using the faint source procedure described in \citet{2021JApA...42...55M} and \citet{Antia2022ApJS..260...40A}, which only considers the intervals in which the background is well behaved. We extract the spectra from layer 1 for both units, as it has the lowest background. The LAXPC spectra are logarithmically re-binned to account for the detector energy resolution and other variability components. 
The spectrum extraction tool in \texttt{LAXPCsoftware22Aug15} also generates a suitable response matrix for each spectrum. The exposure time for both LAXPC units is 40.3~ks with a intensity of $8.27 \pm 0.29$~counts~s$^{-1}$ and $7.33 \pm 0.07$~counts~s$^{-1}$ for LXP10 and LXP20, respectively.

\subsection{{\it MAXI} light curve}\label{MAXI_lightcurve}
\begin{figure*}

          \centering
           \includegraphics[width=1.05\textwidth]{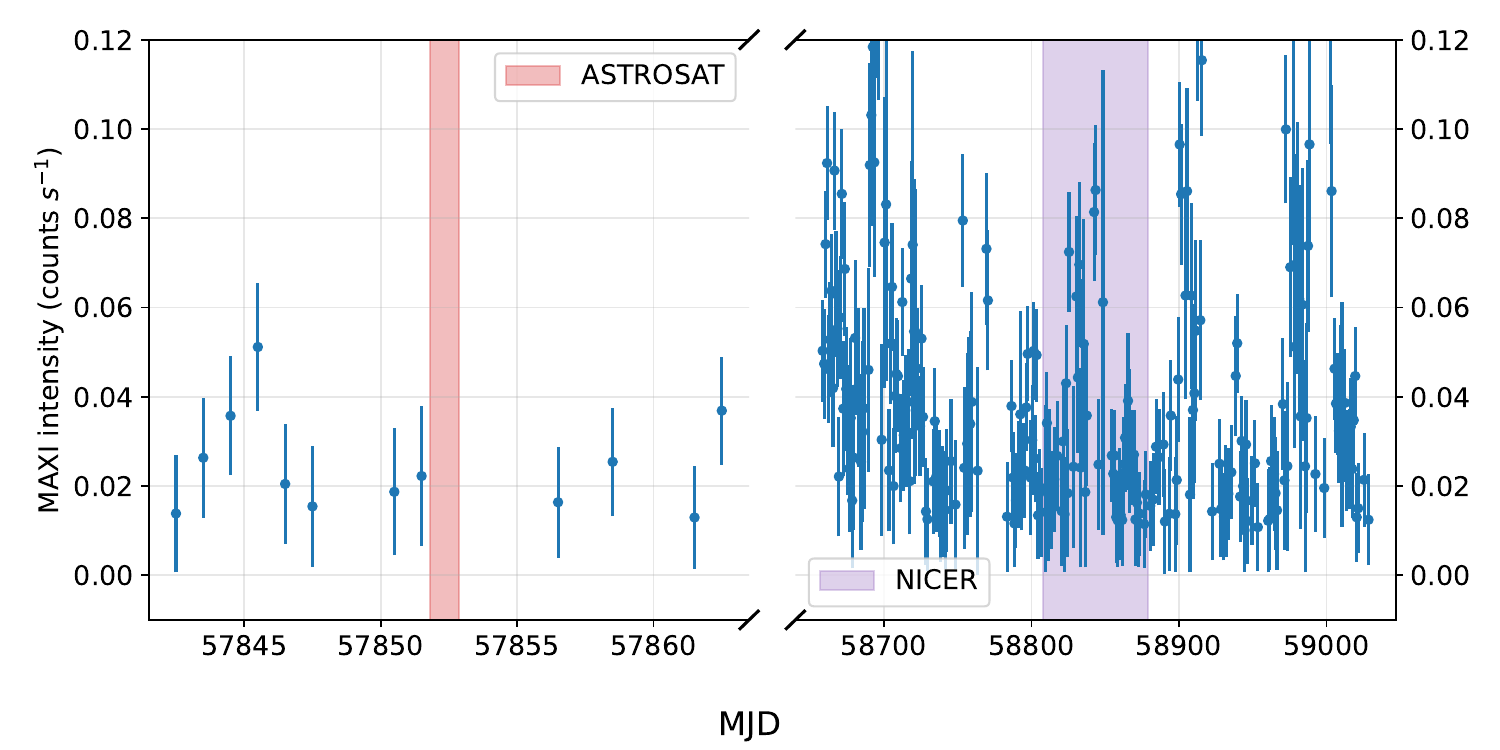}

\caption{Long-term {\it MAXI} light curves of \src\ 
(see Section~\ref{MAXI_lightcurve}). The {\it MAXI} intensities are in 2--20~keV with one-day time bins with $1\sigma$ errors. The broken $x$-axis separates a zoom around the \astr\ epoch (left) from a broader view that includes the \nicer\ campaign (right). Shaded bands mark the \astr\ (red) and \nicer\ (violet) observation windows. This plot places the pointed observations in their long-term context and shows that \src\ remains persistently active with modest variability during both intervals.}\label{fig:MAXI_lc}

\end{figure*}

To investigate the long-term evolution of \src, we refer to the {\it MAXI} \citep{MAXI2009PASJ...61..999M} light curve of the source in the full energy band (2--20~keV) and two of the softer bands (2--4~keV and 4--10~keV). The standard background-subtracted light curves are procured from the {\it MAXI}-Riken website\footnote{\url{http://maxi.riken.jp/star_data/J1249-590/J1249-590.html}}. We depict the full-band light curve in different panels of Figure~\ref{fig:MAXI_lc}. 
Since the source is faint during its persistent state, the standard background-subtracted light curve  may have over-subtraction in some of the energy bands, and thus the time bins where the estimated rate is either negative or consistent with zero within 1$\sigma$ are excluded from the plotting. 

\begin{figure}[!htbp]
    \centering
    \includegraphics[width=1.0\columnwidth]{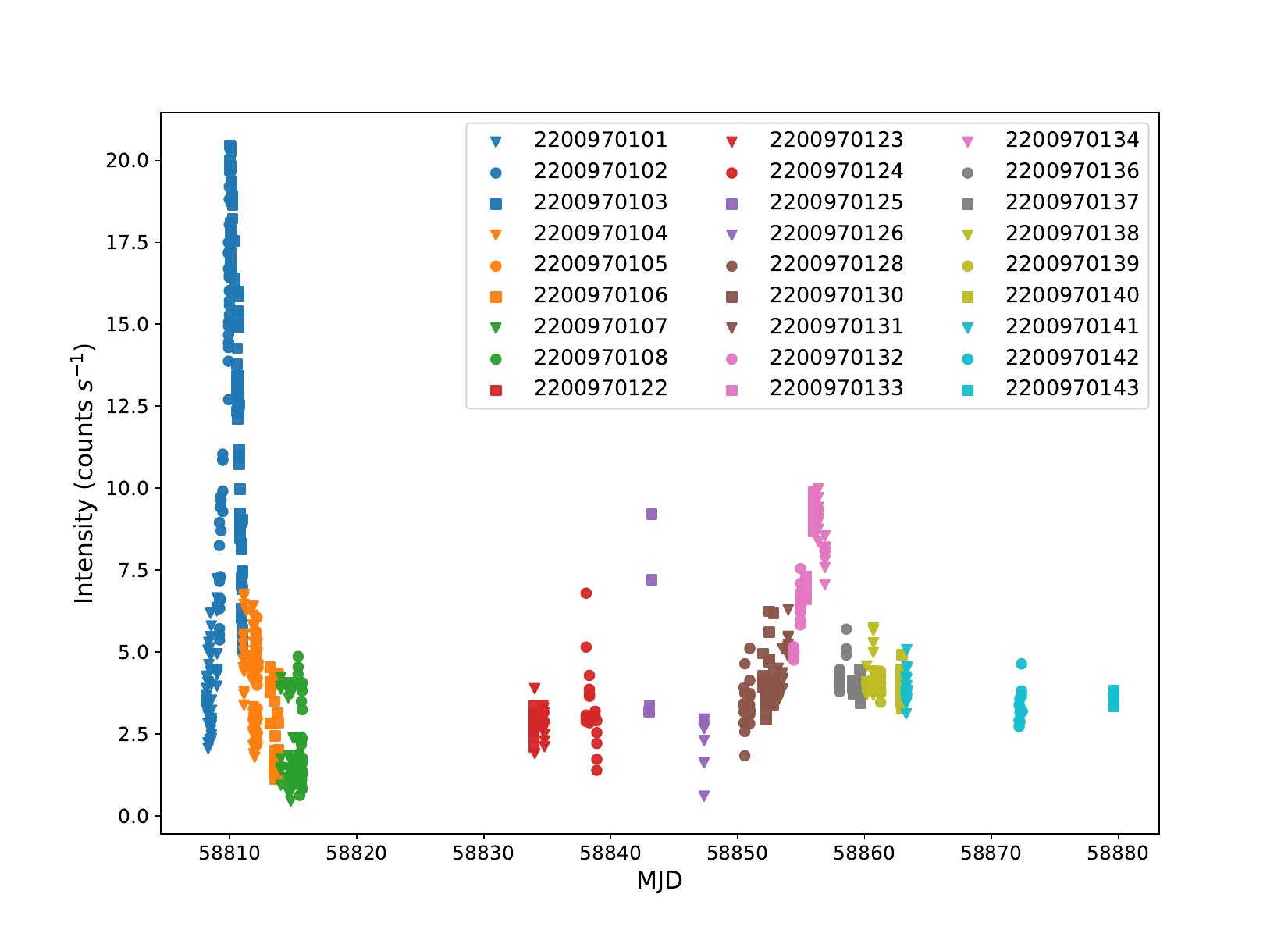}
    \caption{Background-subtracted light curve of \src\ in 3--10~keV as observed by \nicer\ (see Section~\ref{sec:lc_hid}). The observations are binned at 100~s. Each observation is color-coded, as indicated in the legend. The details of the observations are mentioned in the Table~\ref{tab:obs_log}. The source displays pronounced intensity variability, alternating between distinct high and low-flux intervals over the course of the \nicer\ campaign.}
    \label{fig:LC}. 
\end{figure}

\section{Data analysis}\label{sec:analysis}

\subsection{Lightcurve and HID analysis}\label{sec:lc_hid}

\begin{figure}[!htbp]
\centering
% \centering
    \includegraphics[width=1.0\linewidth]{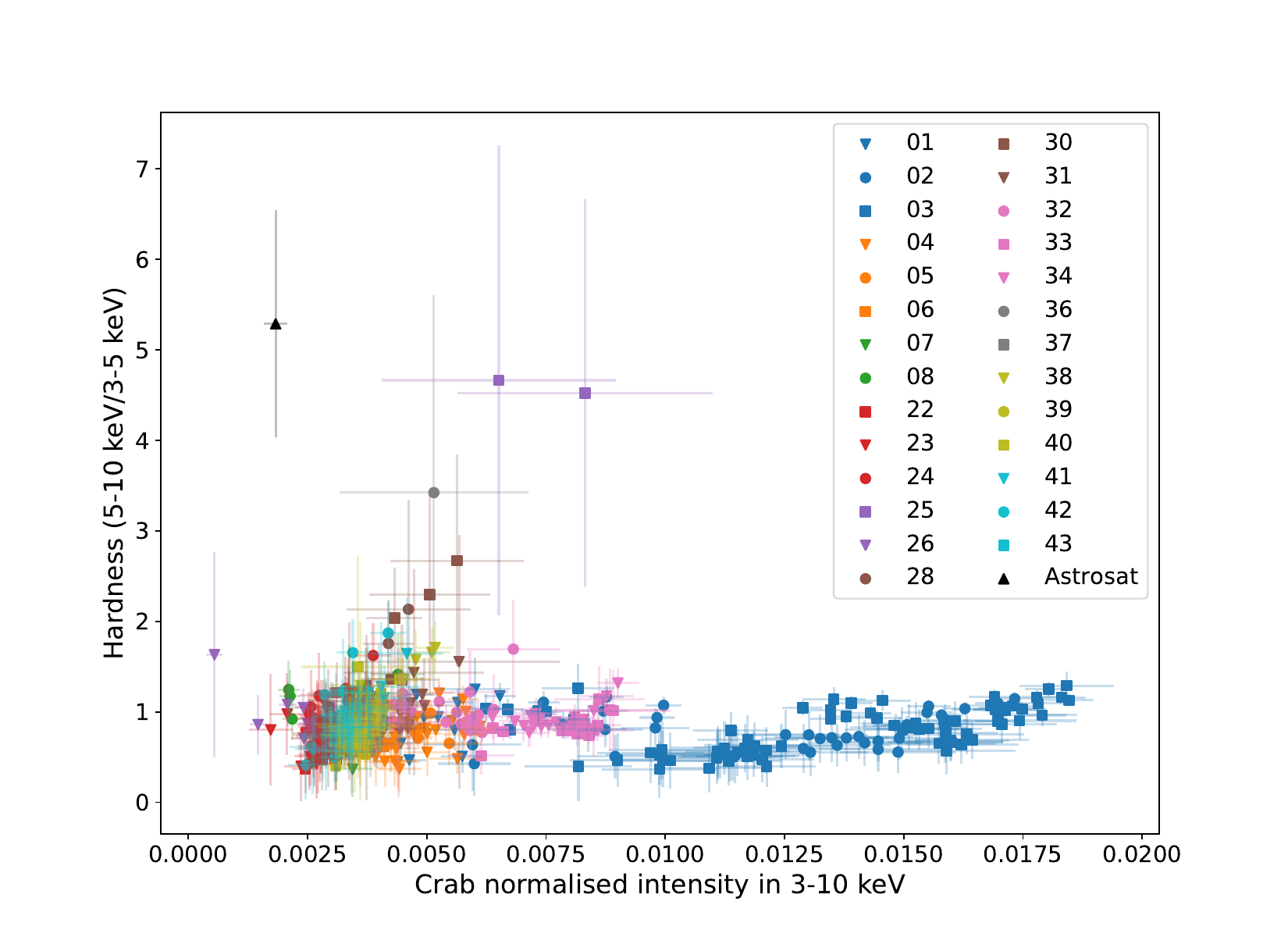} 
\caption{Hardness-intensity diagram (HID) of \src\ using the \astr\ and \nicer\ background-subtracted and Crab-normalized light curves in soft band: 3--5~keV, hard band: 5--10~keV, and full band: 3--10~keV (see section \ref{sec:lc_hid}). The color, marker coding, and the binning for the HID are kept identical to Figure~\ref{fig:LC}.  The epoch of the \astr\ observation is indicated with a black triangle.
In the legend, the two digits correspond to the observation ID 22009701XX to improve the clarity of the plot. This HID highlights two non-simultaneous banana-like tracks at similar hardness but different intensities and a harder, lower intensity island-like region, indicating atoll source behavior and long term accretion variability.} 
    \label{fig:HID}
\end{figure}

To place the pointed observations in a broader flux context, we begin with the long-term \textit{MAXI} light curve (Figure~\ref{fig:MAXI_lc}), which shows modest, recurrent rises in the 2--20~keV daily rate but no coherent periodicity. These variations establish the overall activity level of the source and mark the epochs during which the \nicer\ and \astr\ observations were obtained.
We extract background-subtracted \nicer\ lightcurves with 100~s bins (Figure~\ref{fig:LC}) for all observations. For \astr-LAXPC, the background-subtracted lightcurve and HID are constructed using LAXPC units LXP10 and LXP20. 
To place both instruments on the same hardness-intensity space, we construct HIDs from background subtracted, Crab normalized lightcurves in a soft band of 3--5~keV, a hard band of 5--10~keV, and a full band of 3--10~keV; hardness is defined as (5--10~keV)/(3--5~keV) (Figure~\ref{fig:HID}). 
The \astr\ epoch is indicated by a black triangle. 
Colors and markers are kept identical across Figures~\ref{fig:LC} and \ref{fig:HID}; each ObsID 22009701\textit{XX} is represented by the same symbol/color in both. Throughout the \nicer\ campaign, the source varies strongly on short timescales, alternating between quiescent stretches and brief flares that peak at the highest 3--10~keV intensity observed (Figure~\ref{fig:LC}).
Two bright intervals are particularly prominent; an early phase (e.g., ObsIDs 2200970102--0103; blue symbols) and a later bright phase (e.g., 2200970133--0134; pink symbols), separated by intermediate states with stable, lower intensity. In the HID, both \nicer\ epochs follow a track consistent with the atoll “banana’’ branch \citep{1989A&A...225...79H}. The second bright interval (pink points), corresponding to the second peak in the \nicer\ light curve, largely retraces this shape, showing that similar spectral states are reached at different flux levels. This HID behavior guides our spectral analysis, allowing us to group observations by hardness and select representative intervals for time-resolved spectroscopy (Section~\ref{sec:spectral_analysis}).

\subsection{Spectral analysis}\label{sec:spectral_analysis}

\begin{figure}
     % \begin{subfigure}{0.8\textwidth}
    \centering
    \includegraphics[width=\columnwidth]{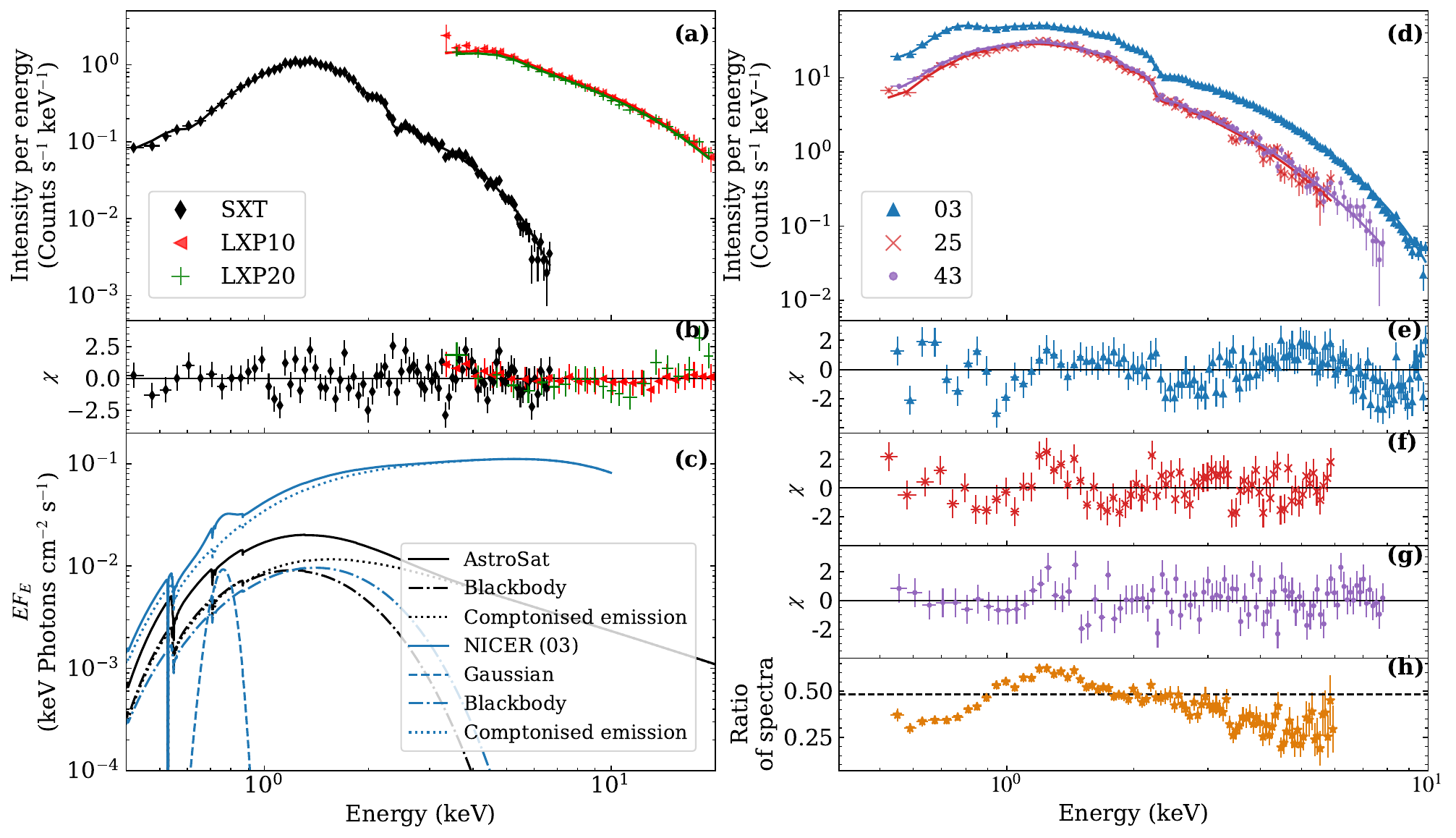} 

\caption{Broadband spectral modeling of \src\ (see Section~\ref{sec:spectral_analysis}). (a) \astr\ SXT+LAXPC count spectra: black diamonds (SXT, 0.4--7~keV), red left-pointing triangles (LXP10, 3--20~keV), and green plus signs (LXP20, 3--20~keV). 
(b) Residuals ($\chi \equiv$~data$-$model over error) for SXT+LAXPC using \texttt{tbabs*(bbodyrad+nthcomp)}. 
(c) Unfolded spectral decomposition for \astr\ and \nicer\ with the \texttt{bbodyrad} and \texttt{nthcomp} components overplotted. 
(d) \nicer\ count spectra (0.5--10~keV); blue upward triangles, red crosses, and purple dots correspond to ObsIDs 22009701\textit{XX} (the legend lists the last two digits \textit{XX}). 
(e)--(g) Residuals ($\chi$) for the \nicer\ spectra in panel (d) using \texttt{tbabs*(gaussian+bbodyrad+nthcomp)}; the \texttt{gaussian} component accounts for the $\sim$0.75~keV instrumental feature. 
(h) Ratio of \nicer\ spectrum \#25 to \#03 illustrating spectral pivoting. 
Together, these panels show that both \astr\ and \nicer\ spectra are well fit with a blackbody plus thermal Comptonization without requiring an additional disk component, and reveal state-dependent spectral changes across the \astr\ and \nicer\ observations.
}

\label{fig:eemodel}

\end{figure}

We analyze the \astr~(SXT+LAXPC) and the \nicer\ spectra of \src\ using the \texttt{xspec} software package (version 12.14.0), adopting $\chi^2$ statistics, and report all confidence intervals at the $1\sigma$ level. 
We generally limit the energy intervals to 0.4--7~keV for the SXT spectrum, 3--20~keV for the LAXPC spectrum, and 0.5--10~keV for the \nicer\ spectra, although some \nicer\ spectra require reduced energy ranges due to stronger background contamination at higher energy ranges.  The SXT spectrum is corrected for gain uncertainties in the response matrix by fitting the gain offset, which is $45 \pm 3~\mathrm{eV}$. 
The observed count spectrum for SXT-LAXPC and \nicer\ are shown in panels (a) and (d) of Figure~\ref{fig:eemodel}, respectively. 
Because the SXT+LAXPC intensity is relatively low and the spectra are not systematics dominated, we do not include any additional systematic uncertainties. Similarly, no systematic uncertainty is applied to the \nicer\ spectra, as the inclusion of the standard recommended systematic error leads to over-fitting and suppresses statistically significant residuals. In most \nicer\ spectra, there is an extra residual at lower energies ($\approx$ 0.75~keV), commonly attributed to calibration uncertainties associated with the oxygen K-edge and the soft X-ray detector response \citep[e.g.][]{Ludlam2018ApJ, Bogdanov2019ApJL}. We model this feature using a (\texttt{gaussian}) component with the centroid energy fixed at 0.75 keV and the width fixed at 0.05 keV, following standard practice in \nicer~analyses \citep[e.g.][]{zhang2023MNRAS.526.3944Z, Putha2024MNRAS.532.3961P}.

We begin with the \nicer\ spectral modeling and then discuss the SXT-LAXPC spectra in that framework. Given that the source is stable within the typical short durations of \nicer\, we model individual observations independently. As a representative case, we describe the spectral modeling of ObsID 2200970103, which has one of the highest intensities (21~counts~s$^{-1}$) and an exposure of 9.16~ks. All spectral models for the \nicer\ observations include a  \texttt{gaussian} component unless otherwise stated (see Table~\ref{tab:NICER_Par_free_e}). Conventionally, spectra of LMXBs require at least two spectral components: a soft thermal component and a hard power-law-like non-thermal component \citep{remillard2006x, Done_2007}. We start by modeling the spectra with an absorbed power-law component (\texttt{tbabs*(gaussian + powerlaw)}), which yields a $\chi^2$ value of 476.92 with 125 degrees of freedom (dof). For modeling the absorption, the abundances are taken from \citet{2000ApJ...542..914W} and the cross-sections from \citet{Vern1996ApJ...465..487V}. To employ a more physical model for the high-energy continuum, we replace the simple power-law component with a thermal Comptonization model (\texttt{nthcomp}; \citealt{1996MNRAS.283..193Z, 1999MNRAS.309..561Z}) which gives a $\chi^2$ value of 243.42 with 124 dof. We next test the spectra with the addition of a blackbody component with the seed photon tied to the blackbody temperature, (\texttt{tbabs*(gaussian + bbodyrad + nthcomp)}) which yields the best fit, with $\chi^2$ = 111.24 for 122 dof. 
Considering that the typical blackbody emission from the source is observed at a temperature of $\gtrsim$1~keV \citep{Zand2008A&A...485..183I} and the typical low-energy rollover in the spectrum is $<$1~keV, we test such temperatures for the spectral fitting. 

\begin{figure}
    \centering  
    \includegraphics[width=0.755\linewidth]{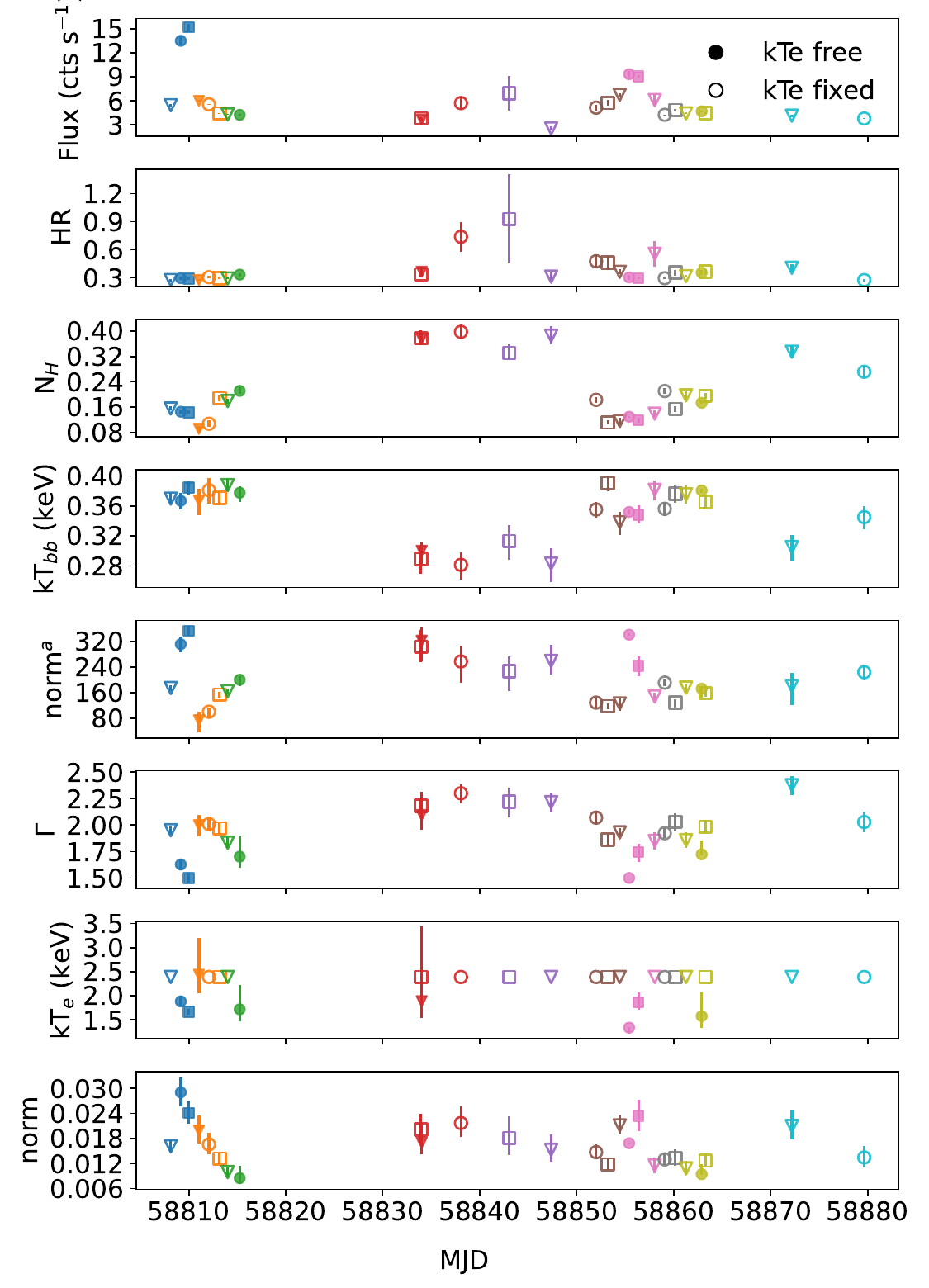} \caption{Temporal evolution of the best-fit spectral parameters obtained from \nicer\ (see Section~\ref{sec:spectral_analysis}) using the best-fit model \texttt{tbabs*(gaussian+bbodyrad+nthcomp)}. Panels from top to bottom show: the 3--10~keV background-subtracted intensity; the hardness ratio (5--10~keV / 3--5~keV); the hydrogen column density $N_{\rm H}$; the blackbody temperature $kT_{\rm bb}$; the \texttt{bbodyrad} normalization (reported as $(R/D_{10})^{2}$); the Comptonization photon index $\Gamma$; the electron temperature $kT_{\rm e}$; and the \texttt{nthcomp} normalization. Each point corresponds to an individual \nicer\ observation (ObsID 22009701\textit{XX}; see Figure~\ref{fig:LC}), color-coded consistently across panels. The hollow symbols indicate observations for which the electron temperature was fixed at $kT_{\rm e}=2.39$~keV. 
    All of the values are as listed in Table~\ref{tab:NICER_Par_free_e}. A narrow \texttt{gaussian} emission component at $\sim$0.75~keV is included when required but is not shown. The evolution reveals clear spectral changes, including correlated variations between the thermal and Comptonized components, indicative of state-dependent spectral evolution.}

    \label{fig:par_evol}
\end{figure}

Introducing a multi-temperature disk component (\texttt{diskbb}) alongside the Comptonized continuum does not improve the fit ($\chi^2$ = 133.37 for 122 dof), and an F-test comparison between models with and without the \texttt{diskbb} component further confirms that a disk contribution is not statistically required. Thus, the \nicer\ spectra favor the {\it Western} configuration (see Section~\ref{sec:intro}), with the soft thermal component arising from the NS surface or boundary layer rather than an optically thick disk. For several \nicer\ observations, the electron temperature is poorly constrained; in these cases, we fix it at 2.39~keV, the value reported by \citet{Zand2008A&A...485..183I}. We plot the unfolded spectral decomposition of the \nicer\ data and the residuals in panels (c),(e), (f) and (g) of Figure~\ref{fig:eemodel}. Table~\ref{tab:NICER_Par_free_e} lists the best-fit parameters obtained with the model \texttt{tbabs*(gaussian+bbodyrad+nthcomp)}, with the electron temperature treated as either free or fixed at 2.39~keV. The evolution of these spectral parameters, which is a major focus of this study, is illustrated in Figure~\ref{fig:par_evol}.

Guided by the \nicer\ results, we then model the \astr\ spectrum using the same physical framework. To account for calibration differences between SXT and LAXPC, we apply an energy-independent scaling factor (\texttt{constant}). Fitting the spectrum with an absorbed power-law model (\texttt{tbabs*powerlaw}) results in a statistically unacceptable fit ($\chi^{2}=418.1$ for 134 dof). Replacing the \texttt{powerlaw} with \texttt{nthcomp} yields a significant improvement ($\chi^{2}$=248.93 for 132 dof). Since the \astr\ spectra show no evidence of a high-energy rollover within the observed 0.4--20~keV band, the electron temperature ($kT_{\rm e}$) is unconstrained. We therefore fix $kT_{\rm e}$ at 100~keV, which renders the Comptonized continuum effectively power-law--like over the AstroSat bandpass; under this assumption, the resulting photon index characterizes the local spectral slope rather than uniquely diagnosing the accretion state. Adding a single-temperature blackbody (\texttt{bbodyrad}) substantially improves it to a $\chi^{2}=139.04$ for 131 dof. Consistent with the \nicer\ observations, adding a multicolor disk component does not improve the fit significantly ($\chi^2$ = 123.17 with 129 dof) and is disfavored by the F‐test, again supporting a Western type of spectral geometry (see Section~\ref{sec:intro}). Thus, we confirm that a single blackbody component adequately describes the soft spectral emission. Hence, our best-fit model for the \astr\ spectrum can be represented as \texttt{constant*tbabs*(bbodyrad+nthcomp)}, with the corresponding best-fit parameters listed in Table~\ref{tab:NICER_Par_free_e}. Panel (c) of Figure~\ref{fig:eemodel} displays the unfolded SXT-LAXPC spectral model, while panel (b) shows the associated residuals.
To ensure a consistent state classification across observations, we base the spectral-state identification on the relative flux contributions of the Comptonized and thermal components in the 0.5--10 keV band of \nicer\ and 0.4--20 keV band of \astr\ (see Table \ref{tab:NICER_Par_free_e}), rather than solely on the photon index $\Gamma$. We define the Comptonized-to-blackbody flux ratio as $R = F_{\rm nthcomp}/F_{\rm bb}$. 
Following the atoll-source framework \citep{1989A&A...225...79H, Lin_2007}, we classify observations with $R \le 6$, corresponding to spectra in which the thermal component contributes significantly or dominates, as the soft state (SS; banana branch). Observations with $R > 6$, indicative of a Comptonization-dominated spectrum, are identified as the hard state (HS; island branch). This quantitative criterion provides a uniform and reproducible demarcation between the two spectral regimes adopted in this work.

\begin{landscape}
%% parameter table NICER
{\scriptsize
\begin{longtable}{l c c c c c c c c c c}
\caption{Best-fit spectral parameters for \src\ from \astr\ (SXT+LAXPC) and \nicer\ (see Section~\ref{sec:spectral_analysis}), using the model  \texttt{tbabs*(gaussian+bbodyrad+nthcomp)} with the Comptonizing electron temperature left free unless noted. 
Units are indicated in the column headers. Errors are $1\sigma$. This table summarizes the spectral decomposition across epochs and enables comparison of the thermal and Comptonized contributions to the observed flux.}

\label{tab:NICER_Par_free_e}
% \begin{longtable}{ccccccccccc}
\\ \toprule \\
 & \texttt{tbabs} & \texttt{gaussian}$^{\dagger}$ & \multicolumn{3}{c}{\texttt{bbodyrad}} & \multicolumn{4}{c}{\texttt{nthcomp}} & \multirow{3}{*}{$\chi^2/\mathrm{dof}$} \\
\cmidrule(lr){2-2}\cmidrule(lr){3-3}\cmidrule(lr){4-6}\cmidrule(lr){7-10}
Obs. ID & $N_{\rm H}$ & norm & $kT_{\rm bb}$ & norm$^{a}$ & Flux$^{c}$ & $\Gamma$ & $kT_{\rm e}$ & norm & Flux$^{c}$ & \\
(State)$^{*}$& ($\times10^{21}$ cm$^{-2}$) & ($\times10^{-3}$) & (keV) & ($\times10^{2}$) & ($\times10^{-11}$)$^{b}$ & & (keV) & ($\times10^{-2}$) & ($\times10^{-11}$)$^{b}$ & \\
\midrule
\endfirsthead

\toprule
\multicolumn{11}{l}{\textit{Table \thetable\ (continued)}}\\
\midrule
\endhead

\midrule
\multicolumn{11}{r}{\textit{Continued on next page}}\\
\endfoot
\bottomrule
\endlastfoot

\multicolumn{11}{c}{\astr$^{d}$} \\
-- &
$2.59^{+0.17}_{-0.18}$ & -- & $0.33 \pm 0.01$ &
$2.47^{+0.29}_{-0.24}$ &
$3.16 \pm 0.11$ &
$2.05 \pm 0.02$ &
$100^{\ddag}$ &
$1.37^{+0.09}_{-0.08}$ &
$11.9 \pm 0.2$ &
$139.04/131$ \\
\midrule

\multicolumn{11}{c}{\nicer} \\
01 (SS) &
$1.56 \pm 0.07$ &
$0.86 \pm 0.09$ &
$0.37 \pm 0.01$ &
$1.74 \pm 0.09$ &
$2.25 \pm 0.07$ &
$1.95 \pm 0.04$ &
$2.39^{e}$ &
$1.62^{+0.14}_{-0.13}$ &
$13.5 \pm 0.3$ &
$151.14/111$ \\

02 (SS) &
$1.46^{+0.07}_{-0.06}$ &
$3.13 \pm 0.14$ &
$0.37 \pm 0.01$ &
$3.12^{+0.23}_{-0.24}$ &
$5.62^{+0.09}_{-0.10}$ &
$1.63^{+0.05}_{-0.06}$ &
$1.88^{+0.12}_{-0.10}$ &
$2.91^{+0.36}_{-0.34}$ &
$30.2 \pm 0.8$ &
$90.49/120$ \\

03 (SS) &
$1.44 \pm 0.05$ &
$3.45 \pm 0.13$ &
$0.38 \pm 0.01$ &
$3.53 \pm 0.16$ &
$7.74^{+0.08}_{-0.09}$ &
$1.50^{+0.05}_{-0.04}$ &
$1.66^{+0.06}_{-0.05}$ &
$2.41^{+0.30}_{-0.25}$ &
$30.5^{+0.8}_{-0.9}$ &
$112.35/123$ \\

04 (HS) &
$0.91^{+0.10}_{-0.09}$ &
$1.47 \pm 0.08$ &
$0.37 \pm 0.02$ &
$0.73^{+0.27}_{-0.38}$ &
$1.31^{+0.07}_{-0.06}$ &
$2.00^{+0.10}_{-0.11}$ &
$2.4^{+0.8}_{-0.4}$ &
$1.99^{+0.36}_{-0.31}$ &
$14.5 \pm 0.2$ &
$100.66/104$ \\

05 (HS) &
$1.08^{+0.12}_{-0.11}$ &
$1.02 \pm 0.10$ &
$0.38 \pm 0.02$ &
$1.00^{+0.14}_{-0.20}$ &
$0.92^{+0.09}_{-0.05}$ &
$2.01 \pm 0.07$ &
$2.39^{e}$ &
$1.66^{+0.29}_{-0.24}$ &
$13.9^{+0.2}_{-0.3}$ &
$85.40/89$ \\

06 (SS) &
$1.88 \pm 0.08$ &
$0.06^{+0.08}_{-0.06}$ &
$0.37 \pm 0.01$ &
$1.54 \pm 0.09$ &
$2.55^{+0.06}_{-0.07}$ &
$1.97 \pm 0.05$ &
$2.39^{e}$ &
$1.32^{+0.14}_{-0.13}$ &
$10.4 \pm 0.4$ &
$124.84/92$ \\

07 (SS) &
$1.79 \pm 0.08$ &
$0.08 \pm 0.08$ &
$0.39 \pm 0.01$ &
$1.64 \pm 0.07$ &
$2.94 \pm 0.06$ &
$1.83 \pm 0.05$ &
$2.39^{e}$ &
$1.00^{+0.11}_{-0.10}$ &
$10.0^{+0.3}_{-0.2}$ &
$121.55/101$ \\

08 (SS) &
$2.1 \pm 0.1$ &
N &
$0.38 \pm 0.01$ &
$2.0^{+0.1}_{-0.2}$ &
$4.07^{+0.06}_{-0.05}$ &
$1.70^{+0.20}_{-0.10}$ &
$1.7^{+0.5}_{-0.2}$ &
$0.85^{+0.29}_{-0.14}$ &
$7.9^{+0.6}_{-0.5}$ &
$84.60/91$ \\

22 (HS) &
$3.77^{+0.25}_{-0.23}$ &
N &
$0.29 \pm 0.02$ &
$3.03^{+0.53}_{-0.48}$ &
$1.46^{+0.20}_{-0.18}$ &
$2.18^{+0.07}_{-0.08}$ &
$2.39^{e}$ &
$2.02^{+0.37}_{-0.31}$ &
$10.8^{+0.7}_{-0.8}$ &
$109.78/78$ \\

23 (SS) &
$3.75^{+0.18}_{-0.13}$ &
N &
$0.30^{+0.01}_{-0.02}$ &
$3.23^{+0.41}_{-0.62}$ &
$2.47 \pm 0.10$ &
$2.09^{+0.22}_{-0.14}$ &
$1.9^{+1.6}_{-0.4}$ &
$1.70^{+0.46}_{-0.28}$ &
$8.9^{+0.7}_{-0.7}$ &
$116.77/88$ \\

24 (SS) &
$4.0 \pm 0.2$ &
N &
$0.28 \pm 0.02$ &
$2.58^{+0.51}_{-0.66}$ &
$3.79 \pm 0.11$ &
$2.30^{+0.08}_{-0.09}$ &
$2.39^{e}$ &
$2.17^{+0.40}_{-0.34}$ &
$6.5^{+0.8}_{-0.5}$ &
$108.74/65$ \\

25 (SS) &
$3.31^{+0.26}_{-0.23}$ &
N &
$0.31^{+0.02}_{-0.03}$ &
$2.27^{+0.47}_{-0.63}$ &
$2.20 \pm 0.18$ &
$2.22^{+0.13}_{-0.15}$ &
$2.39^{e}$ &
$1.81^{+0.52}_{-0.41}$ &
$9.2 \pm 1.0$ &
$93.78/71$ \\

26 (HS)&
$3.9 \pm 0.3$ &
N &
$0.28 \pm 0.02$ &
$2.6^{+0.5}_{-0.4}$ &
$0.84^{+0.16}_{-0.07}$ &
$2.22 \pm 0.09$ &
$2.39^{e}$ &
$1.53^{+0.36}_{-0.28}$ &
$8.3^{+0.4}_{-0.6}$ &
$82.94/87$ \\

28 (SS) &
$2.71^{+0.12}_{-0.11}$ &
N &
$0.33^{+0.01}_{-0.02}$ &
$2.29^{+0.06}_{-0.68}$ &
$2.77^{+0.07}_{-0.08}$ &
$1.95^{+0.38}_{-0.05}$ &
$1.6^{+9.8}_{-0.4}$ &
$1.3^{+0.5}_{-0.1}$ &
$8.2^{+0.6}_{-0.3}$ &
$103.66/76$ \\

30 (SS) &
$1.83 \pm 0.09$ &
$0.28 \pm 0.07$ &
$0.36 \pm 0.01$ &
$1.28^{+0.13}_{-0.15}$ &
$5.31^{+0.03}_{-0.02}$ &
$2.07 \pm 0.06$ &
$2.39^{e}$ &
$1.47^{+0.18}_{-0.16}$ &
$5.6 \pm 0.1$ &
$114.82/77$ \\

31 (SS) &
$1.12^{+0.08}_{-0.07}$ &
$0.76 \pm 0.07$ &
$0.36 \pm 0.01$ &
$0.9 \pm 0.1$ &
$2.11^{+0.05}_{-0.07}$ &
$2.00 \pm 0.06$ &
$2.39^{e}$ &
$2.12^{+0.25}_{-0.21}$ &
$11.4 \pm 0.3$ &
$87.84/87$ \\

32 (HS) &
$1.52^{+0.09}_{-0.08}$ &
$1.73 \pm 0.10$ &
$0.36 \pm 0.01$ &
$1.09^{+0.13}_{-0.12}$ &
$0.91 \pm 0.09$ &
$2.03 \pm 0.06$ &
$2.39^{e}$ &
$1.84^{+0.23}_{-0.20}$ &
$16.1 \pm 0.3$ &
$82.02/97$ \\

33 (SS) &
$1.30 \pm 0.03$ &
$2.80 \pm 0.13$ &
$0.35 \pm 0.01$ &
$3.41 \pm 0.04$ &
$5.17 \pm 0.04$ &
$1.50 \pm 0.01$ &
$1.33^{+0.01}_{-0.12}$ &
$1.69 \pm 0.01$ &
$17.1 \pm 0.1$ &
$109.34/88$ \\

34 (SS) &
$1.19^{+0.09}_{-0.08}$ &
$2.49 \pm 0.12$ &
$0.35 \pm 0.01$ &
$2.44^{+0.28}_{-0.33}$ &
$3.53^{+0.09}_{-0.08}$ &
$1.74^{+0.08}_{-0.09}$ &
$1.87^{+0.21}_{-0.17}$ &
$2.35^{+0.38}_{-0.36}$ &
$19.9^{+0.6}_{-0.7}$ &
$126.56/108$ \\

36 (SS) &
$1.4 \pm 0.1$ &
$0.4 \pm 0.1$ &
$0.38 \pm 0.01$ &
$1.47^{+0.11}_{-0.12}$ &
$2.65^{+0.09}_{-0.08}$ &
$1.85 \pm 0.08$ &
$2.39^{e}$ &
$1.2 \pm 0.2$ &
$10.8^{+0.5}_{-0.6}$ &
$104.02/84$ \\

37 (SS) &
$2.1 \pm 0.1$ &
N &
$0.36 \pm 0.01$ &
$1.91 \pm 0.12$ &
$2.68^{+0.08}_{-0.07}$ &
$1.92 \pm 0.05$ &
$2.39^{e}$ &
$1.30^{+0.14}_{-0.13}$ &
$10.3 \pm 0.5$ &
$120.81/102$ \\

38 (SS) &
$1.54 \pm 0.09$ &
$0.37 \pm 0.08$ &
$0.38 \pm 0.01$ &
$1.28^{+0.12}_{-0.14}$ &
$5.96^{+0.03}_{-0.04}$ &
$2.03 \pm 0.08$ &
$2.39^{e}$ &
$1.3 \pm 0.2$ &
$5.4 \pm 0.1$ &
$101.34/78$ \\

39 (SS) &
$1.97^{+0.12}_{-0.11}$ &
$0.31 \pm 0.12$ &
$0.38 \pm 0.01$ &
$1.75^{+0.13}_{-0.12}$ &
$2.85 \pm 0.09$ &
$1.86 \pm 0.07$ &
$2.39^{e}$ &
$1.08^{+0.17}_{-0.15}$ &
$10.1^{+0.5}_{-0.6}$ &
$105.98/97$ \\

40 (SS) &
$1.74 \pm 0.03$ &
$0.41 \pm 0.07$ &
$0.38 \pm 0.01$ &
$1.72^{+0.02}_{-0.28}$ &
$3.62^{+0.03}_{-0.02}$ &
$1.72^{+0.13}_{-0.01}$ &
$1.58^{+0.49}_{-0.24}$ &
$0.95^{+0.24}_{-0.01}$ &
$8.3 \pm 0.1$ &
$114.85/91$ \\

41 (SS) &
$1.96 \pm 0.09$ &
$0.16 \pm 0.08$ &
$0.37 \pm 0.01$ &
$1.58 \pm 0.11$ &
$2.80^{+0.07}_{-0.06}$ &
$1.98 \pm 0.06$ &
$2.39^{e}$ &
$1.27^{+0.16}_{-0.15}$ &
$9.3 \pm 0.4$ &
$120.12/89$ \\

42 (SS) &
$3.35^{+0.20}_{-0.18}$ &
N &
$0.31 \pm 0.02$ &
$1.81^{+0.42}_{-0.60}$ &
$4.47^{+0.11}_{-0.10}$ &
$2.38^{+0.08}_{-0.09}$ &
$2.39^{e}$ &
$2.10^{+0.40}_{-0.33}$ &
$5.9^{+0.2}_{-0.6}$ &
$88.57/72$ \\

43 (SS) &
$2.72^{+0.18}_{-0.17}$ &
N &
$0.35 \pm 0.02$ &
$2.24^{+0.24}_{-0.23}$ &
$2.46 \pm 0.13$ &
$2.03 \pm 0.10$ &
$2.39^{e}$ &
$1.35^{+0.28}_{-0.24}$ &
$9.8^{+0.7}_{-0.8}$ &
$95.39/87$ \\

\end{longtable}
}

\begin{flushleft}
\footnotesize
$^{*}$ HS = hard state, SS = soft state. \\
$^{\dagger}$ Line energy and width for the Gaussian component are frozen at 0.75~keV and 0.05~keV, respectively.\\
$^{\ddag}$ Parameter frozen.\\
$^{a}$ \texttt{bbodyrad} normalization: ${\rm norm}=(R_{\rm bb}/D_{10})^2$, where $R_{\rm bb}$ is the apparent emitting radius (km) and $D_{10}$ is distance in units of 10~kpc. The apparent radius inferred from the \texttt{bbodyrad} normalization does not include spectral hardening or gravitational redshift effects, and the physical emitting radius can be estimated as $R \simeq f_{\rm c}^2 R_{\rm bb}/(1+z)$, where $f_{\rm c}$ is the 
color-correction factor and $z$ is the gravitational redshift.\\
$^{b}$ Unabsorbed flux in erg~cm$^{-2}$~s$^{-1}$ estimated with \texttt{cflux} on the corresponding component.\\
$^{c}$ Flux is over 0.5--10~keV for \nicer\ and 0.4--20~keV for \astr.\\
$^{d}$ Gain offset $=45\pm3$~eV; cross-normalization constants $C_{\rm LXP10}=0.85\pm0.03$, $C_{\rm LXP20}=0.88\pm0.02$; $C_{\rm SXT}=1$ (fixed).\\
$^{e}$ indicates the parameter was fixed at that value. \\
N: not detected. 
\end{flushleft}
\end{landscape}

\section{Results and Discussion}\label{sec:res}

 We investigate the variability of the UCXB \src\ by combining long-term {\it MAXI}/GSC monitoring with pointed \nicer\ and \astr\ observations. The {\it MAXI} light curve reveals long-term flux variations that contextualize the shorter timescale variability observed with \nicer. Broadband spectral modeling of the \nicer\ and \astr\ data corroborates earlier results \citep{Zand2008A&A...485..183I} and enables a systematic study of changes in the relative contributions of thermal and Comptonized emission across luminosity states. The \astr\ observation, obtained during a low-intensity phase, further probes the source at harder energies (0.4--20~keV), offering additional constraints on the thermal and Comptonized components. Together, these observations present the first coherent view of the spectral evolution of \src\ across distinct luminosity states, highlighting its variable accretion behavior. 
 
\subsection{Light curve properties using \it{NICER} and \it{MAXI}}\label{sec:LC}

We characterize the X-ray variability of \src\ on day timescales using pointed \nicer\ observations and long-term monitoring data from \textit{MAXI}. The \nicer\ observations (Figure~\ref{fig:LC}) samples the source densely over $\sim$75~days and shows pronounced intensity variations, including distinct rises in the 3--10~keV intensity. These variations are accompanied by correlated changes in hardness (see Section~\ref{sec:lc_hid}), indicating that the observed variability arises from spectral-state transitions, likely driven by fluctuations in the accretion rate, which govern the changing balance between the thermal and Comptonized emission components \citep{Done_2007}. On longer timescales, \citet{Zand2008A&A...485..183I} reported variability ranging from weeks to hundreds of days based on ASM/PCA/BAT monitoring, highlighting a persistently active source. To place the \nicer\ coverage in this broader context, we depict the \textit{MAXI} 1-day light curve spanning several months around the \nicer\ and \textit{AstroSat} epochs. Over these baselines, the \textit{MAXI} light curve traces the long-term flux evolution and confirms that the \nicer\ observation occurred during an active phase. For the \astr\ epoch, \textit{MAXI} coverage is sparse, and the surrounding flux measurements are low, consistent with the soft, low-intensity state we identify in our spectral analysis.
The epochs of the \nicer\ and \astr\ observations fall within activity levels consistent with the high and low-flux states reported by \citet{Zand2008A&A...485..183I}.
Based on long-term \textit{MAXI} monitoring, \citet{asai2022PASJ...74..974A} classified \src\ as a modified periodic (MP) source exhibiting quasi-periodic variations on $\sim$2.5--10~yr timescales. In the \citet{asai2022PASJ...74..974A} light curve, the epoch corresponding to our \astr\ observation lies in a relatively low-flux interval, whereas the \nicer\ campaign overlaps a brighter phase of the long-term modulation. This matches our MAXI snapshots, which show low daily rates around the \astr\ window and significantly elevated activity during the \nicer\ window. 
This indicates that \src\ exhibits multi-timescale variability: short-term flux changes captured by the pointed observations are superposed on longer-term luminosity modulations reported in previous studies \citep{Barret2000, asai2022PASJ...74..974A}. 
Finally, the \nicer\ campaign (0.2--12~keV) reinforces the picture established by \citet{Zand2008A&A...485..183I} of \src\ as a persistently active source exhibiting moderate, short-term flux variability. These variations likely arise from short-term spectral-state changes driven by fluctuations in the accretion flow, reflecting evolving accretion and Comptonization conditions (see Sections~\ref{sec:lc_hid} and \ref{sec:spectral_analysis}).

\subsection{HID properties using {\it NICER} and {\it AstroSat}}

The HID (Figure~\ref{fig:HID}) provides a clear view of the source’s spectral-state behavior across the \nicer\ and \astr\ observations. The \astr\ epoch (black point; Figure~\ref{fig:HID}) occupies a region of relatively low intensity and high hardness. This comparatively harder location of the \astr\ point in the HID arises from the adopted hardness definition (5--10 keV / 3--5 keV), which probes energy bands where the Comptonized component dominates even in the soft state. Consequently, the hardness ratio appears elevated relative to NICER-based HIDs that include softer energy bands more sensitive to the thermal emission. $R = F_{\rm nthcomp}/F_{\rm bb} \approx 3.8$. 
According to our adopted flux-ratio criterion ($R \le 6$ for the soft state), this places the source in the soft state. Although the \astr\ point appears on the harder side of the HID, this location reflects the chosen hardness definition (5--10 keV / 3--5 keV), which emphasizes the Comptonized emission. The flux decomposition confirms that the thermal component still contributes substantially to the total emission (see Table \ref{tab:NICER_Par_free_e}).

Within the \nicer\ data the HID reveals a horizontal track (blue points; Figure~\ref{fig:HID}) resembling the “banana” branch commonly observed in atoll sources \citep[e.g.,][]{1989A&A...225...79H,2004astro.ph.10551V}. At a later epoch, the source traces a second set of points (pink; Figure~\ref{fig:HID}) at lower intensity, representing the early phase of a transition toward the softer banana branch. These two excursions of a banana-like track, while slightly varying in peak flux, share comparable hardness and position on the HID. This suggests that the source reached similar spectral states across distinct epochs, likely reflecting moderate long-term variations in the accretion rate. Additionally, between the epochs tracing the higher-intensity banana track (blue points; Figure~\ref{fig:HID}) and those characterized by low intensity and low hardness, some observations extend toward higher hardness at similarly low intensities. Among these, ObsID~2200970125 (purple; Figure~\ref{fig:HID}) occupies a region of low intensity and high hardness, consistent with island-state morphology \citep{1989A&A...225...79H,Gierlinski2002}.
These HID patterns are consistent with the classifications observed in other atoll sources \citep[e.g.,][]{2014MNRAS.438.2784C}. This state is generally associated with lower accretion rates and enhanced Comptonization, in contrast to the banana branch, where thermal emission dominates. In the context of state transitions, the pink points likely represent the progression of the Comptonization dominated island state towards the thermally dominated banana branch, with the spectrum softening as the thermal component becomes more prominent.

Overall, the HID demonstrates that \src\ samples multiple spectral states, with the banana branch showing short-term spectral state changes reflecting evolving accretion and Comptonization, and the island state representing a harder, low-luminosity regime. In this context, spectral state identification is based on hardness and relative component fluxes rather than on the photon index alone, which can be model-dependent when the Comptonizing electron temperature is fixed or unconstrained \citep{Lin_2007}.
 These spectral-state mappings, when combined with the temporal information from the light curves (Section~\ref{sec:LC}), provide a comprehensive view of the source’s accretion behavior and state transitions.

\subsection{Evolution of the spectral properties of \src}\label{sec:evol_spectral_properties}
 
Motivated by the source’s evolution along the HID, we investigate its spectral properties during the persistent state to assess how they vary across different spectral states.
The observed broadband flux of \src\ in 0.5--10~keV of \nicer\ data (see Figure~\ref{fig:LC}) is between $0.7\times 10^{-10}$ and $3.7\times10^{-10}\,\mathrm{erg\,cm^{-2}\,s^{-1}}$.
The epochs analyzed here are at flux levels comparable to the PCA low-flux state reported by \citet{Zand2008A&A...485..183I}. The source exhibits a broadband spectrum that requires multiple components to be adequately modeled. The dominant emission components in the 0.5--10~keV \nicer\ band, and the 0.4--7~keV \astr-SXT and 3--20~keV \astr-LAXPC band are shown in Figure~\ref{fig:eemodel}. Another aspect to emphasize is \astr's broader energy coverage compared to \nicer, which enables tighter constraints on spectral components {\citep{singh2016SPIE.9905E..1ES, ArmasPadilla2023A&A}}.

\section{Spectral Evolution and Accretion States}\label{sec:evol_spectral_&_accretion}

Our spectral analysis reveals that the absorption column density to sufficiently model the soft X-ray spectra of \src\ lies within $1\times10^{21}$~cm$^{-2}$ to $4\times10^{21}$~cm$^{-2}$ for \nicer\ observations. This is broadly consistent with the line of sight column density as measured from the HI maps \citep[3.35$\times10^{21}$~cm$^{-2}$;][]{HI4PI2016A&A...594A.116H} and also consistent with the archival observations \citep{Zand2008A&A...485..183I}. However, the relatively wide spread in fitted $N_{\mathrm{H}}$ values across individual \nicer\ spectra, which is larger than typically expected for ultra-compact systems, does not necessarily imply genuine variability in the absorbing material and may instead reflect degeneracies between $N_{\mathrm{H}}$ and other spectral components in the modeling. The analysis of the \astr\ observations also require a $N_{\mathrm{H}} = 2.6\times10^{21}$~cm$^{-2}$ which is consistent with the previous estimates. Overall, the fitted column densities are consistent with previous measurements, and the available data do not provide clear evidence for long-term changes in the absorption toward the source.

Typically in NS UCXBs, the thermal emission is modeled as originating either from the NS boundary layer or from the accretion disk. In the present analysis, where the thermal component is represented solely by a blackbody, the inferred blackbody temperature spans 0.28–0.39~keV. The blackbody normalization (see Table~\ref{tab:NICER_Par_free_e} for definition) lies in the range 100–400, corresponding to an apparent emitting radius of 
$R_{\rm bb} \simeq 4.3$--8.6~km. This represents the apparent radius derived from 
the \texttt{bbodyrad} model and does not include spectral hardening effects. 
Applying a representative color-correction factor 
$f_{\rm c} \simeq 1.6$, appropriate for neutron-star surface or boundary-layer 
emission \citep[e.g.,][]{Suleimanov2012}, and neglecting gravitational redshift for 
simplicity, the inferred physical radius increases to 
$R \simeq 6.9$--13.8~km. These values remain consistent with emission from an extended region at the disk–NS interface, such as a boundary layer, rather than from a localized hot spot on the stellar surface alone \citep{Inogamov_1999, Lin_2007}. The inclusion of an additional multicolor disk blackbody component does not result in a statistically significant improvement in the fits, as confirmed by the F-test. Given the ultra-compact nature of the system and its short orbital period, the accretion disk is expected to be physically small (e.g., \citealt{Frank2002, vanHaaften2012}). In such a configuration, any emission from a cool, compact disk would be intrinsically weak and readily outshone by the Comptonized component fed by seed photons from the NS surface. The absence of a detectable disk blackbody contribution in our fits is therefore consistent with Comptonization occurring in an extended accretion-disk corona, as described in the Birmingham (Western) model geometry for LMXBs, where the spectrum is dominated by a soft blackbody boundary layer and a hard Comptonized component \citep{Church2004}.
The low-flux spectrum of \src\ from \textit{RXTE}/PCA in \citet{Zand2008A&A...485..183I} indicates that the spectrum has to be modeled with a power-law. \citet{Zand2008A&A...485..183I} also find that this power-law is better described by a Comptonized emission, which in our analysis is \texttt{nthcomp}. For several \nicer\ observations, the electron temperature could not be independently constrained due to the limited energy coverage. In these cases, we fixed the electron temperature at $kT_{\rm e} = 2.39$~keV, adopting the value reported for the high-flux Comptonized component in the \textit{Swift--XRT+PCA} analysis of \citet{Zand2008A&A...485..183I}. Fixing $kT_{\rm e}$ stabilizes the fits and allows the photon index ($\Gamma$) to be constrained; under this assumption, small variations in $\Gamma$ likely reflect parameter degeneracy rather than intrinsic changes in the Comptonizing plasma.
In the \nicer\ observations, although the 0.5--10~keV flux is often dominated by the Comptonized component, the fitted spectral parameters ($\Gamma$ and $kT_{\rm e}$), together with the inferred optical depth ($\tau$), indicate a soft, optically thick Comptonization regime \citep{1996MNRAS.283..193Z, Done_2007}(see Table~\ref{tab:NICER_Par_free_e}).  
In the \astr\ observation, no high-energy cutoff is detected; therefore, the Comptonized component is constrained by the photon index, and the electron temperature is fixed at $kT_{\rm e}=100$~keV to approximate a power-law spectrum over the observed bandpass \citep{2017ApJ...850..155S}.
The resulting photon index, $\Gamma \simeq 2.05$, characterizes the spectral slope under this assumption. While this value lies within the range of photon indices measured during the softer, banana-like states in the \nicer\ campaign, the \nicer\ spectra require a much lower electron temperature ($kT_{\rm e}\simeq 2$–3~keV), implying a physically distinct Comptonizing plasma. Because $\Gamma$ in the \texttt{nthcomp} model depends explicitly on $kT_{\rm e}$, the Comptonization parameters derived from the \astr\ and \nicer\ spectra are not directly comparable, and similarities in $\Gamma$ do not imply comparable coronal properties. The apparent differences in electron temperature are therefore most naturally attributed to the differing energy coverage and instrumental responses of \astr\ and \nicer, which are known to introduce systematic differences in inferred Comptonization parameters \citep{Kohlemainen2014, Madsen2017AJ}. In addition, the source occupies different regions of the hardness–intensity diagram during the \astr\ and \nicer\ epochs, and the Comptonization parameters, particularly the electron temperature; differ markedly, together suggesting that the source was observed in different spectral states, consistent with the long-term flux differences shown in Figure~\ref{fig:MAXI_lc}. As noted above, the photon index $\Gamma$ is not used as a primary spectral-state diagnostic in this work, because its inferred value depends sensitively on the assumed electron temperature and on the limited energy coverage of the instruments, particularly for NICER. Instead, spectral states are identified using the HID and the relative flux contributions of the Comptonized and thermal components, which provide a more robust and physically meaningful classification across both epochs.

Building on these considerations, we examine the evolution of the Comptonized component in relation to the thermal emission using the \texttt{nthcomp} model. Across the \nicer\ observations, we find a clear inverse relationship between the photon index ($\Gamma$) and the blackbody temperature (Figure~\ref{fig:par_evol}). Over the observed range of blackbody temperatures ($kT_{\rm bb} \simeq 0.27$--0.39~keV), the characteristic energy of the seed photons increases correspondingly, shifting the low-energy rollover of the Comptonized spectrum toward higher energies \citep{Done_2007}.
Over the observed blackbody temperature range, the photon index spans $\Gamma \simeq 1.5$--2.4, with harder spectra generally associated with higher blackbody temperatures, indicating a broad anti-correlation between $kT_{\rm bb}$ and $\Gamma$. Restricting the analysis to observations in which the electron
temperature is free, a Spearman rank correlation test yields $r_s = -0.6$ with a p-value of $8.8\times10^{-2}$ ($n = 9$). While the inferred
trend is consistent with an anti-correlation, its statistical significance is
marginal given the limited sample size.
This behavior is consistent with the source transitioning between island-like and banana-like accretion states. Harder spectra, with photon indices $\Gamma \sim 1.5$, are typically associated with island states characterized by reduced seed photon input and a relatively stronger Comptonized contribution, whereas softer spectra ($\Gamma \sim 2.4$) correspond to banana states with higher accretion rates and enhanced thermal emission from the disk or boundary layer \citep{Barret2000, ArmasPadilla2017MNRAS, 1996MNRAS.283..193Z}.
However, the substantial scatter in $\Gamma$, particularly in observations where the electron temperature is fixed at 2.39~keV, suggests that part of this trend is influenced by parameter coupling in the spectral modeling rather than reflecting purely intrinsic changes in the Comptonizing plasma.

Using the photon indices and electron plasma temperature, the optical depth of the Comptonizing cloud ($\tau$) can be estimated following \citet{1980A&A....86..121S}: \[
\Gamma = -\frac{1}{2} + \sqrt{\frac{9}{4} + \gamma}, \quad \text{where} \quad \gamma = \pi^3 \frac{kT_e}{m_e c^2} \left(\tau^2 + \frac{2}{3} \right)
\]. This yields optical depths in the range $\tau \simeq 3.3$–6.7, consistent with a soft or banana-state Comptonization regime \citep{Gierlinski2002, Done_2007}. We note that these values depend on the assumed electron temperature; with the current data, the degeneracy between $kT_e$ and $\tau$ cannot be resolved. In the subset of nine observations where $kT_{\rm e}$ is left free, the electron temperature spans $kT_{\rm e}\simeq1.34$–2.44~keV. While these variations broadly coincide with changes in source hardness and intensity, the limited high-energy sensitivity of the data prevents a systematic assessment of $kT_{\rm e}$ across all observations.

\begin{figure}
     
    \centering
    \includegraphics[width=1\columnwidth]{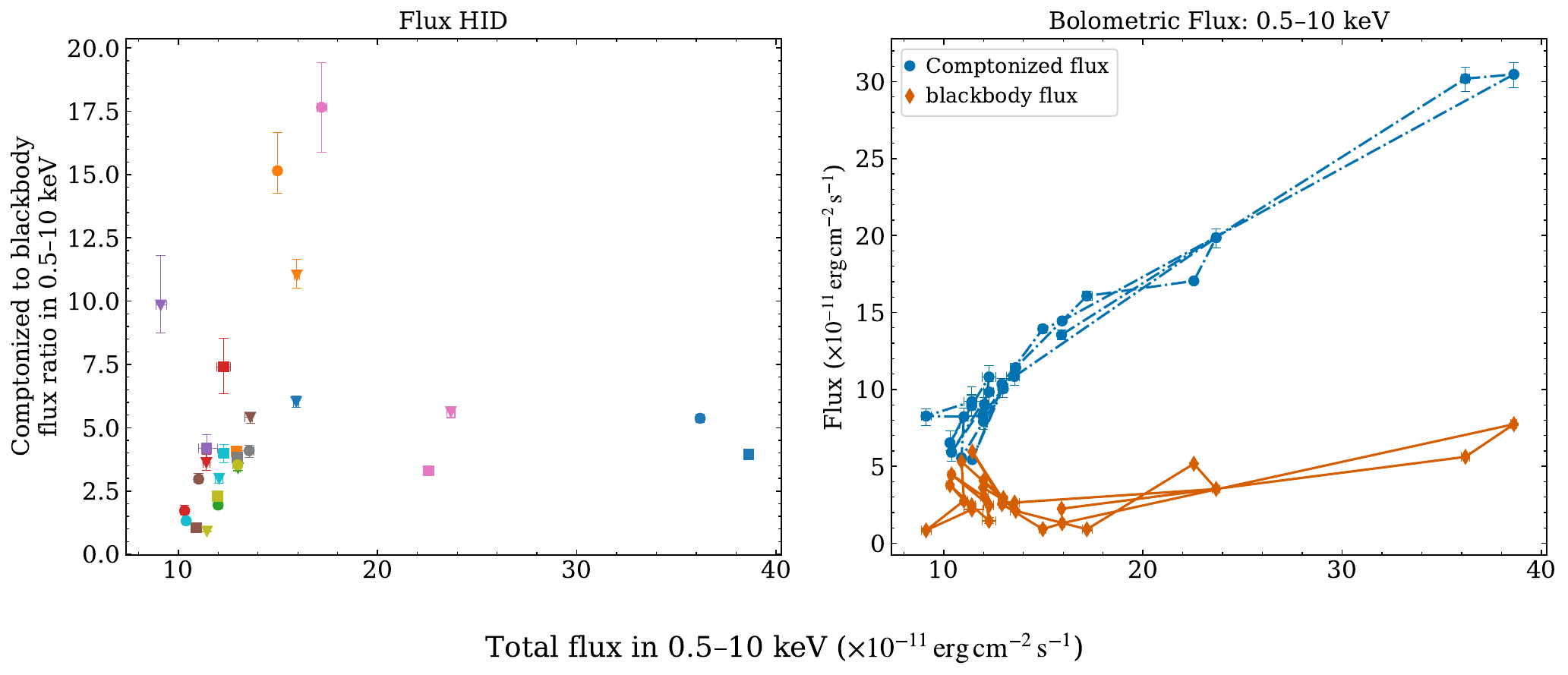} 
    
\caption{Flux ratios of all \nicer\ observations (see section \ref{sec:evol_spectral_&_accretion}). Left Panel: Ratio of the Comptonized flux to blackbody flux as a function of the total flux in 0.5--10~keV. The color and symbols of the observations are kept identical to Figure \ref{fig:LC}. This panel demonstrates consistency with the HID. Right Panel: Evolution of the Comptonized flux and the blackbody flux as a function of the total flux in 0.5--10~keV as well. This panel highlights a clear anti-correlation between the thermal and Comptonized components across the observed flux range. 
}\label{fig:fluxratio}
\end{figure}

To investigate the origin of the spectral evolution, Figure~\ref{fig:fluxratio} illustrates how the relative contributions of the Comptonized and thermal components govern the \nicer\ spectra. In the left panel of Figure~\ref{fig:fluxratio}, we plot the ratio of the Comptonized flux to the blackbody flux as a function of the total 0.5–10~keV flux. At lower total fluxes, corresponding to island states, the flux ratio is high, indicating that the Comptonized component dominates the emission. As the source brightens, this ratio decreases systematically, reflecting the growing contribution of the blackbody component. This behavior qualitatively reproduces the hardness–intensity evolution seen in the HID (Figure~\ref{fig:HID}), demonstrating that changes in spectral hardness arise from variations in the relative strengths of the Comptonized and thermal components. The right panel of Figure~\ref{fig:fluxratio} shows the evolution of the individual component fluxes with total flux. The Comptonized and blackbody fluxes exhibit a clear anti-correlation, with the emission at lower luminosities dominated by the Comptonized component, corresponding to island-like behavior. Above this range, the increasing blackbody flux reflects a growing thermal contribution to the total emission. For the \astr\ observation, the Comptonized-to-blackbody flux ratio in the 0.4--20 keV band is 
$R \approx 3.8$ (see Table \ref{tab:NICER_Par_free_e}), placing it within the soft-state regime under our adopted criterion. 
This indicates that, despite a significant Comptonized contribution, the thermal component remains energetically important. In the context of accretion geometry, such spectra are consistent with configurations in which the optically thick inner accretion flow contributes prominently, while Comptonization remains active but does not dominate the energy budget.
 A direct illustration of the state-dependent spectral changes in the \nicer\ observations is provided by the comparison between ObsID~200970103 and ObsID~200970125, shown in panel (h) of Figure~\ref{fig:eemodel}. ObsID~03, which lies on the banana branch of the HID, exhibits a softer spectrum characterized by a higher blackbody temperature ($kT_{\rm bb}\simeq0.38$~keV), a larger blackbody normalization (corresponding to a larger emitting area), indicative of enhanced thermal emission from the NS boundary layer and a relatively weaker Comptonized contribution. In contrast, ObsID~25, associated with the island state, shows a harder spectrum with a lower blackbody temperature ($kT_{\rm bb}\simeq0.31$~keV), and reduced blackbody flux, reflecting a diminished thermal component and a comparatively stronger role of Comptonization (Table~\ref{tab:NICER_Par_free_e}). This ratio spectrum clearly demonstrates spectral pivoting, with the island-state spectrum becoming harder at higher energies while being relatively suppressed at soft energies, in agreement with the flux–ratio trends. While such behavior is qualitatively consistent with changes in the balance between the thermal and Comptonized emission components as the source transitions between the banana and island states, we note that significant degeneracy exists between $N_{\mathrm{H}}$, the soft thermal component ($kT_{\rm bb}$), and the Comptonized continuum. Consequently, the relative flux contributions of individual components are not uniquely constrained. The observed spectral evolution therefore likely reflects changes in the overall spectral shape, which may arise from variations in the relative strengths of the thermal and Comptonized emission, as expected for atoll-type NS systems \citep{1989A&A...225...79H, Gierlinski2002}.

\section{Summary}

We present a comprehensive temporal and spectral study of the ultra-compact X-ray binary \src\ using long-term \textit{MAXI}/GSC monitoring, and pointed observations from \textit{NICER} and \textit{AstroSat}. Our analysis establishes, for the first time, a coherent picture of the source’s spectral evolution across different flux states.

\begin{itemize}
\item Our \astr\ observation falls in a low-flux segment of the MAXI light-curve variations, while the \nicer\ campaign coincides with a brighter phase. Within these intervals, \nicer\ data reveals strong short-term flux excursions atop a broadly stable emission level.

\item The \nicer\ HID shows an atoll-like pattern, with a single banana-like track, along which the source undergoes two phases of higher intensity and a cluster of hard, low-flux points marking an island state, indicating transitions between Comptonization and thermal dominance as the source evolves in luminosity.

\item The emission consists of a soft blackbody component ($kT = 0.27$--$0.39$~keV, radius $\sim$7--14~km) consistent with boundary-layer emission, plus a harder Comptonized tail that steepens from $\Gamma \approx 1.5$--2.4 as the source brightens.
    
\item The spectral evolution is driven by shifts in the balance between the boundary-layer emission and the Comptonizing corona, which vary in a contrasting manner across the 0.5--10 keV flux range, with the transition from island to banana states corresponding to a softening of the spectrum as the thermal component strengthens and the Comptonized emission weakens.

\end{itemize}

%The last numbered section should briefly summarise what has been done, and describe the final conclusions which the authors draw from their work.

%% Also note that the akcnowlodgment environment does not support long amounts of text. If you have a lot of people and institutions to acknowledge, do not use this command. Instead, create a new \section{Acknowledgments}.
\section*{Acknowledgments}
\noindent 
This research has made use of \nicer’s data obtained through the High Energy Astrophysics Science Archive Research Center (HEASARC)
Online Service, provided by the NASA/Goddard Space Flight Center.
This work has also utilized data from the {\it AstroSat} mission of the Indian Space Research Organisation (ISRO), archived at the Indian Space Science Data Centre (ISSDC). 
The paper has used data from the SXT and the LAXPC instruments developed at the Tata Institute of Fundamental Research (TIFR), Mumbai. The {\it AstroSat} Payload Operation Centers at TIFR are thanked for verifying and releasing the data via the ISSDC data archive and providing the necessary software tools.

%This research has also made use of data and/or software provided by the High Energy Astrophysics Science Archive Research Center (HEASARC), which is a service of the Astrophysics Science Division at NASA/GSFC and the High Energy Astrophysics Division of the Smithsonian Astrophysical Observatory. 
% \end{acknowledgments}


\vspace{5mm}
\textbf{Facilities}: \astr, \nicer


\textbf{Software}: {astropy \citep{2018AJ....156..123A},  
        % matplotlib \rough{add cite}
          }
\vspace{5mm}

\noindent       
\textbf{Declaration of generative AI and AI-assisted technologies in the manuscript preparation process}

\vspace{5mm}
\noindent 
During the preparation of this work the author(s) used ChatGPT (OpenAI) in order to assist with grammar checking and sentence refinement. After using this tool/service, the author(s) reviewed and edited the content as needed and take(s) full responsibility for the content of the published article.

\bibliography{ref}  % <- your .bib file

@ARTICLE{2000ApJ...542..914W,
       author = {{Wilms}, J. and {Allen}, A. and {McCray}, R.},
        title = "{On the Absorption of X-Rays in the Interstellar Medium}",
      journal = {Astrophys. J.},
         year = 2000,
        month = oct,
       volume = {542},
       number = {2},
        pages = {914-924},
          doi = {10.1086/317016},
archivePrefix = {arXiv},
       eprint = {astro-ph/0008425},
 primaryClass = {astro-ph},
       adsurl = {https://ui.adsabs.harvard.edu/abs/2000ApJ...542..914W}
}

@ARTICLE{Vern1996ApJ...465..487V,
   author = {{Verner}, D.~A. and {Ferland}, G.~J. and {Korista}, K.~T. and 
	{Yakovlev}, D.~G.},
    title = "{Atomic Data for Astrophysics. II. New Analytic FITS for Photoionization Cross Sections of Atoms and Ions}",
  journal = {Astrophys. J.},
   eprint = {astro-ph/9601009},
     year = 1996,
    month = jul,
   volume = 465,
    pages = {487},
      doi = {10.1086/177435},
   adsurl = {http://adsabs.harvard.edu/abs/1996ApJ...465..487V}
}

@ARTICLE{2021JApA...42...55M,
       author = {{Misra}, Ranjeev and {Roy}, Jayashree and {Yadav}, J.~S.},
        title = "{An alternative scheme to estimate AstroSat/LAXPC background for faint sources}",
      journal = {Journal of Astrophysics and Astronomy},
         year = 2021,
        month = oct,
       volume = {42},
       number = {2},
          eid = {55},
        pages = {55},
          doi = {10.1007/s12036-021-09734-2},
archivePrefix = {arXiv},
       eprint = {2102.06402},
 primaryClass = {astro-ph.IM},
       adsurl = {https://ui.adsabs.harvard.edu/abs/2021JApA...42...55M}
}

@ARTICLE{Inogamov_1999,
       author = {{Inogamov}, N.~A. and {Sunyaev}, R.~A.},
        title = "{Spread of matter over a neutron-star surface during disk accretion}",
      journal = {Astronomy Letters},
         year = 1999,
        month = may,
       volume = {25},
       number = {5},
        pages = {269-293},
          doi = {10.48550/arXiv.astro-ph/9904333},
archivePrefix = {arXiv},
       eprint = {astro-ph/9904333},
 primaryClass = {astro-ph},
       adsurl = {https://ui.adsabs.harvard.edu/abs/1999AstL...25..269I}
}

@ARTICLE{Antia2022ApJS..260...40A,
       author = {{Antia}, H.~M. and {Agrawal}, P.~C. and {Katoch}, Tilak and {Manchanda}, R.~K. and {Mukerjee}, Kallol and {Shah}, Parag},
        title = "{Improved Background Model for the Large Area X-Ray Proportional Counter (LAXPC) Instrument on board AstroSat}",
      journal = {Astrophys. J. Suppl. Ser.},
         year = 2022,
        month = jun,
       volume = {260},
       number = {2},
          eid = {40},
        pages = {40},
          doi = {10.3847/1538-4365/ac6dd0},
archivePrefix = {arXiv},
       eprint = {2205.03136},
 primaryClass = {astro-ph.IM},
       adsurl = {https://ui.adsabs.harvard.edu/abs/2022ApJS..260...40A}
}

@ARTICLE{Yadav2017CSci..113..591Y,
       author = {{Yadav}, J.~S. and {Agrawal}, P.~C. and {Antia}, H.~M. and {Manchanda}, R.~K. and {Paul}, B. and {Misra}, Ranjeev},
        title = "{Large Area X-ray Proportional Counter instrument on AstroSat}",
      journal = {Current Science},
         year = 2017,
        month = aug,
       volume = {113},
       number = {4},
        pages = {591},
          doi = {10.18520/cs/v113/i04/591-594},
archivePrefix = {arXiv},
       eprint = {1705.06440},
 primaryClass = {astro-ph.IM},
       adsurl = {https://ui.adsabs.harvard.edu/abs/2017CSci..113..591Y}
}

@ARTICLE{Singh2017JApA...38...29S,
       author = {{Singh}, K.~P. and {Stewart}, G.~C. and {Westergaard}, N.~J. and {Bhattacharayya}, S. and {Chandra}, S. and {Chitnis}, V.~R. and {Dewangan}, G.~C. and {Kothare}, A.~T. and {Mirza}, I.~M. and {Mukerjee}, K. and {Navalkar}, V. and {Shah}, H. and {Abbey}, A.~F. and {Beardmore}, A.~P. and {Kotak}, S. and {Kamble}, N. and {Vishwakarama}, S. and {Pathare}, D.~P. and {Risbud}, V.~M. and {Koyande}, J.~P. and {Stevenson}, T. and {Bicknell}, C. and {Crawford}, T. and {Hansford}, G. and {Peters}, G. and {Sykes}, J. and {Agarwal}, P. and {Sebastian}, M. and {Rajarajan}, A. and {Nagesh}, G. and {Narendra}, S. and {Ramesh}, M. and {Rai}, R. and {Navalgund}, K.~H. and {Sarma}, K.~S. and {Pandiyan}, R. and {Subbarao}, K. and {Gupta}, T. and {Thakkar}, N. and {Singh}, A.~K. and {Bajpai}, A.},
        title = "{Soft X-ray Focusing Telescope Aboard AstroSat: Design, Characteristics and Performance}",
      journal = {Journal of Astrophysics and Astronomy},
         year = 2017,
        month = jun,
       volume = {38},
       number = {2},
          eid = {29},
        pages = {29},
          doi = {10.1007/s12036-017-9448-7},
       adsurl = {https://ui.adsabs.harvard.edu/abs/2017JApA...38...29S}
}

@INPROCEEDINGS{Yadav2016SPIE.9905E..1DY,
       author = {{Yadav}, J.~S. and {Agrawal}, P.~C. and {Antia}, H.~M. and {Chauhan}, Jai Verdhan and {Dedhia}, Dhiraj and {Katoch}, Tilak and {Madhwani}, P. and {Manchanda}, R.~K. and {Misra}, Ranjeev and {Pahari}, Mayukh and {Paul}, B. and {Shah}, Parag},
        title = "{Large Area X-ray Proportional Counter (LAXPC) instrument onboard ASTROSAT}",
    booktitle = {Space Telescopes and Instrumentation 2016: Ultraviolet to Gamma Ray},
         year = 2016,
       editor = {{den Herder}, Jan-Willem A. and {Takahashi}, Tadayuki and {Bautz}, Marshall},
       series = {Proc. SPIE},
       volume = {9905},
        month = jul,
          eid = {99051D},
        pages = {99051D},
          doi = {10.1117/12.2231857},
       adsurl = {https://ui.adsabs.harvard.edu/abs/2016SPIE.9905E..1DY}
}

@INPROCEEDINGS{Singh2014SPIE.9144E..1SS,
       author = {{Singh}, Kulinder Pal and {Tandon}, S.~N. and {Agrawal}, P.~C. and {Antia}, H.~M. and {Manchanda}, R.~K. and {Yadav}, J.~S. and {Seetha}, S. and {Ramadevi}, M.~C. and {Rao}, A.~R. and {Bhattacharya}, D. and {Paul}, B. and {Sreekumar}, P. and {Bhattacharyya}, S. and {Stewart}, G.~C. and {Hutchings}, J. and {Annapurni}, S.~A. and {Ghosh}, S.~K. and {Murthy}, J. and {Pati}, A. and {Rao}, N.~K. and {Stalin}, C.~S. and {Girish}, V. and {Sankarasubramanian}, K. and {Vadawale}, S. and {Bhalerao}, V.~B. and {Dewangan}, G.~C. and {Dedhia}, D.~K. and {Hingar}, M.~K. and {Katoch}, T.~B. and {Kothare}, A.~T. and {Mirza}, I. and {Mukerjee}, K. and {Shah}, H. and {Shah}, P. and {Mohan}, R. and {Sangal}, A.~K. and {Nagabhusana}, S. and {Sriram}, S. and {Malkar}, J.~P. and {Sreekumar}, S. and {Abbey}, A.~F. and {Hansford}, G.~M. and {Beardmore}, A.~P. and {Sharma}, M.~R. and {Murthy}, S. and {Kulkarni}, R. and {Meena}, G. and {Babu}, V.~C. and {Postma}, J.},
        title = "{ASTROSAT mission}",
    booktitle = {Space Telescopes and Instrumentation 2014: Ultraviolet to Gamma Ray},
         year = 2014,
       editor = {{Takahashi}, Tadayuki and {den Herder}, Jan-Willem A. and {Bautz}, Mark},
       series = {Proc. SPIE},
       volume = {9144},
        month = jul,
          eid = {91441S},
        pages = {91441S},
          doi = {10.1117/12.2062667},
       adsurl = {https://ui.adsabs.harvard.edu/abs/2014SPIE.9144E..1SS}
}

@INPROCEEDINGS{singh2016SPIE.9905E..1ES,
       author = {{Singh}, Kulinder Pal and {Stewart}, Gordon C. and {Chandra}, Sunil and {Mukerjee}, Kallol and {Kotak}, Sanket and {Beardmore}, Andy P. and {Chitnis}, Varsha and {Dewangan}, Gulab C. and {Bhattacharyya}, Sudip and {Mirza}, Irfan and {Kamble}, Nilima and {Navalkar}, Vinita and {Shah}, Harshit and {Vishwakarma}, S. and {Koyande}, J.},
        title = "{In-orbit performance of SXT aboard AstroSat}",
    booktitle = {Space Telescopes and Instrumentation 2016: Ultraviolet to Gamma Ray},
         year = 2016,
       editor = {{den Herder}, Jan-Willem A. and {Takahashi}, Tadayuki and {Bautz}, Marshall},
       series = {Proc. SPIE},
       volume = {9905},
        month = jul,
          eid = {99051E},
        pages = {99051E},
          doi = {10.1117/12.2235309},
       adsurl = {https://ui.adsabs.harvard.edu/abs/2016SPIE.9905E..1ES}
}

@ARTICLE{Bassa2006A&A...446L..17B,
       author = {{Bassa}, C.~G. and {Jonker}, P.~G. and {in't Zand}, J.~J.~M. and {Verbunt}, F.},
        title = "{Two new candidate ultra-compact X-ray binaries}",
      journal = {Astron. Astrophys.},
         year = 2006,
        month = feb,
       volume = {446},
       number = {3},
        pages = {L17-L20},
          doi = {10.1051/0004-6361:200500229},
archivePrefix = {arXiv},
       eprint = {astro-ph/0601045},
 primaryClass = {astro-ph},
       adsurl = {https://ui.adsabs.harvard.edu/abs/2006A&A...446L..17B}
}

@ARTICLE{jonker2007MNRAS.378.1187J,
       author = {{Jonker}, P.~G. and {in't Zand}, J.~J.~M. and {M{\'e}ndez}, M. and {van der Klis}, M.},
        title = "{Detection of a 1258-Hz high-amplitude kilohertz quasi-periodic oscillation in the ultracompact X-ray binary 1A 1246-588}",
      journal = {Mon. Not. R. Astron. Soc.},
         year = 2007,
        month = jul,
       volume = {378},
       number = {3},
        pages = {1187-1190},
          doi = {10.1111/j.1365-2966.2007.11854.x},
archivePrefix = {arXiv},
       eprint = {0704.1741},
 primaryClass = {astro-ph},
       adsurl = {https://ui.adsabs.harvard.edu/abs/2007MNRAS.378.1187J}
}

@ARTICLE{Zand2008A&A...485..183I,
       author = {{in't Zand}, J.~J.~M. and {Bassa}, C.~G. and {Jonker}, P.~G. and {Keek}, L. and {Verbunt}, F. and {M{\'e}ndez}, M. and {Markwardt}, C.~B.},
        title = "{An X-ray and optical study of the ultracompact X-ray binary A 1246-58}",
      journal = {Astron. Astrophys.},
         year = 2008,
        month = jul,
       volume = {485},
       number = {1},
        pages = {183-194},
          doi = {10.1051/0004-6361:200809361},
archivePrefix = {arXiv},
       eprint = {0804.2666},
 primaryClass = {astro-ph},
       adsurl = {https://ui.adsabs.harvard.edu/abs/2008A&A...485..183I}
}

@ARTICLE{piro1997IAUC.6538....2P,
       author = {{Piro}, L. and {Heise}, J. and {Jager}, R. and {Feroci}, M. and {D'Andreta}, G. and {Spoliti}, G. and {Coletta}, A. and {Muller}, H. and {Ricci}, D.},
        title = "{New X-Ray Burster}",
      journal = {IAU Circ.},
         year = 1997,
        month = jan,
       volume = {6538},
        pages = {2},
       adsurl = {https://ui.adsabs.harvard.edu/abs/1997IAUC.6538....2P}
}

@ARTICLE{boller1997IAUC.6546....1B,
       author = {{Boller}, T. and {Haberl}, F. and {Voges}, W. and {Piro}, L. and {Heise}, J.},
        title = "{New X-Ray Burster}",
      journal = {IAU Circ.},
         year = 1997,
        month = jan,
       volume = {6546},
        pages = {1},
       adsurl = {https://ui.adsabs.harvard.edu/abs/1997IAUC.6546....1B}
}

@ARTICLE{Doesburgh2018MNRAS.479..426V,
       author = {{van Doesburgh}, Marieke and {van der Klis}, Michiel and {Morsink}, Sharon M.},
        title = "{The highest frequency kHz QPOs in neutron star low-mass X-ray binaries}",
      journal = {Mon. Not. R. Astron. Soc.},
         year = 2018,
        month = sep,
       volume = {479},
       number = {1},
        pages = {426-434},
          doi = {10.1093/mnras/sty1404},
archivePrefix = {arXiv},
       eprint = {1805.11361},
 primaryClass = {astro-ph.HE},
       adsurl = {https://ui.adsabs.harvard.edu/abs/2018MNRAS.479..426V}
}

@ARTICLE{carpenter1977MNRAS.179P..27C,
       author = {{Carpenter}, G.~F. and {Eyles}, C.~J. and {Skinner}, G.~K. and {Wilson}, A.~M. and {Willmore}, A.~P.},
        title = "{New cosmic X-ray sources observed by the RMC experiment on Ariel V.}",
      journal = {Mon. Not. R. Astron. Soc.},
         year = 1977,
        month = apr,
       volume = {179},
        pages = {27P-34},
          doi = {10.1093/mnras/179.1.27P},
       adsurl = {https://ui.adsabs.harvard.edu/abs/1977MNRAS.179P..27C}
}

@ARTICLE{Barret2000,
       author = {{Barret}, D. and {Olive}, J.~F. and {Boirin}, L. and {Done}, C. and {Skinner}, G.~K. and {Grindlay}, J.~E.},
        title = "{Hard X-Ray Emission from Low-Mass X-Ray Binaries}",
      journal = {Astrophys. J.},
         year = 2000,
        month = apr,
       volume = {533},
       number = {1},
        pages = {329-351},
          doi = {10.1086/308651},
archivePrefix = {arXiv},
       eprint = {astro-ph/9911042},
 primaryClass = {astro-ph},
       adsurl = {https://ui.adsabs.harvard.edu/abs/2000ApJ...533..329B}
}

@article{ArmasPadilla2017MNRAS,
  author = {Armas Padilla, M. and Degenaar, N. and Wijnands, R.},
  title = {X-ray spectral states of accreting neutron stars},
  journal = {Mon. Not. R. Astron. Soc.},
  volume = {467},
  number = {3},
  pages = {290--304},
  year = {2017},
  doi = {10.1093/mnras/stx079}
}

@article{ArmasPadilla2023A&A,
	author = {{Armas Padilla, M.} and {Corral-Santana, J. M.} and {Borghese, A.} and {Cúneo, V. A.} and {Muñoz-Darias, T.} and {Casares, J.} and {Torres, M. A. P.}},
	title = {UltraCompCAT: A comprehensive catalogue of ultra-compact and short orbital period X-ray binaries},
	DOI= "10.1051/0004-6361/202346797",
	journal = {Astron. Astrophys.},
	year = 2023,
	volume = 677,
	pages = "A186",
}

@ARTICLE{asai2022PASJ...74..974A,
       author = {{Asai}, Kazumi and {Mihara}, Tatehiro and {Matsuoka}, Masaru},
        title = "{Decades-long variations in NS-LMXBs observed with MAXI/GSC, RXTE/ASM, and Ginga/ASM}",
      journal = {Publ. Astron. Soc. Jpn.},
         year = 2022,
        month = aug,
       volume = {74},
       number = {4},
        pages = {974-990},
          doi = {10.1093/pasj/psac049},
archivePrefix = {arXiv},
       eprint = {2206.02299},
 primaryClass = {astro-ph.HE},
       adsurl = {https://ui.adsabs.harvard.edu/abs/2022PASJ...74..974A}
}

@ARTICLE{bhattacharya1991PhR...203....1B,
       author = {{Bhattacharya}, D. and {van den Heuvel}, E.~P.~J.},
        title = "{Formation and evolution of binary and millisecond radio pulsars}",
      journal = {Phys. Rep.},
         year = 1991,
        month = jan,
       volume = {203},
       number = {1-2},
        pages = {1-124},
          doi = {10.1016/0370-1573(91)90064-S},
       adsurl = {https://ui.adsabs.harvard.edu/abs/1991PhR...203....1B}
}

@ARTICLE{MAXI2009PASJ...61..999M,
       author = {{Matsuoka}, Masaru and {Kawasaki}, Kazuyoshi and {Ueno}, Shiro and {Tomida}, Hiroshi and {Kohama}, Mitsuhiro and {Suzuki}, Motoko and {Adachi}, Yasuki and {Ishikawa}, Masaki and {Mihara}, Tatehiro and {Sugizaki}, Mutsumi and {Isobe}, Naoki and {Nakagawa}, Yujin and {Tsunemi}, Hiroshi and {Miyata}, Emi and {Kawai}, Nobuyuki and {Kataoka}, Jun and {Morii}, Mikio and {Yoshida}, Atsumasa and {Negoro}, Hitoshi and {Nakajima}, Motoki and {Ueda}, Yoshihiro and {Chujo}, Hirotaka and {Yamaoka}, Kazutaka and {Yamazaki}, Osamu and {Nakahira}, Satoshi and {You}, Tetsuya and {Ishiwata}, Ryoji and {Miyoshi}, Sho and {Eguchi}, Satoshi and {Hiroi}, Kazuo and {Katayama}, Haruyoshi and {Ebisawa}, Ken},
        title = "{The MAXI Mission on the ISS: Science and Instruments for Monitoring All-Sky X-Ray Images}",
      journal = {Publ. Astron. Soc. Jpn.},
         year = 2009,
        month = oct,
       volume = {61},
        pages = {999},
          doi = {10.1093/pasj/61.5.999},
archivePrefix = {arXiv},
       eprint = {0906.0631},
 primaryClass = {astro-ph.IM},
       adsurl = {https://ui.adsabs.harvard.edu/abs/2009PASJ...61..999M}
}

@ARTICLE{2018AJ....156..123A,
       author = {{Astropy Collaboration} and {Price-Whelan}, A.~M. and {Sip{\H{o}}cz}, B.~M. and {G{\"u}nther}, H.~M. and {Lim}, P.~L. and {Crawford}, S.~M. and {Conseil}, S. and {Shupe}, D.~L. and {Craig}, M.~W. and {Dencheva}, N. and {Ginsburg}, A. and {VanderPlas}, J.~T. and {Bradley}, L.~D. and {Astropy Contributors}},
        title = "{The Astropy Project: Building an Open-science Project and Status of the v2.0 Core Package}",
      journal = {Astron. J.},
         year = 2018,
        month = sep,
       volume = {156},
       number = {3},
          eid = {123},
        pages = {123},
          doi = {10.3847/1538-3881/aabc4f},
archivePrefix = {arXiv},
       eprint = {1801.02634},
 primaryClass = {astro-ph.IM},
       adsurl = {https://ui.adsabs.harvard.edu/abs/2018AJ....156..123A}
}

@ARTICLE{zhang2023MNRAS.526.3944Z,
       author = {{Zhang}, Liang and {M{\'e}ndez}, Mariano and {Garc{\'\i}a}, Federico and {Zhang}, Yuexin and {Ma}, Ruican and {Altamirano}, Diego and {Yang}, Zi-Xu and {Ma}, Xiang and {Tao}, Lian and {Huang}, Yue and {Jia}, Shumei and {Zhang}, Shuang-Nan and {Qu}, Jinlu and {Song}, Liming and {Zhang}, Shu},
        title = "{Type-A quasi-periodic oscillation in the black hole transient MAXI J1348-630}",
      journal = {Mon. Not. R. Astron. Soc.},
         year = 2023,
        month = dec,
       volume = {526},
       number = {3},
        pages = {3944-3950},
          doi = {10.1093/mnras/stad3062},
archivePrefix = {arXiv},
       eprint = {2310.04208},
 primaryClass = {astro-ph.HE},
       adsurl = {https://ui.adsabs.harvard.edu/abs/2023MNRAS.526.3944Z}
}

@INPROCEEDINGS{2014SPIE.9144E..20A,
       author = {{Arzoumanian}, Z. and {Gendreau}, K.~C. and {Baker}, C.~L. and {Cazeau}, T. and {Hestnes}, P. and {Kellogg}, J.~W. and {Kenyon}, S.~J. and {Kozon}, R.~P. and {Liu}, K.-C. and {Manthripragada}, S.~S. and {Markwardt}, C.~B. and {Mitchell}, A.~L. and {Mitchell}, J.~W. and {Monroe}, C.~A. and {Okajima}, T. and {Pollard}, S.~E. and {Powers}, D.~F. and {Savadkin}, B.~J. and {Winternitz}, L.~B. and {Chen}, P.~T. and {Wright}, M.~R. and {Foster}, R. and {Prigozhin}, G. and {Remillard}, R. and {Doty}, J.},
        title = "{The neutron star interior composition explorer (NICER): mission definition}",
    booktitle = {Space Telescopes and Instrumentation 2014: Ultraviolet to Gamma Ray},
         year = 2014,
       editor = {{Takahashi}, Tadayuki and {den Herder}, Jan-Willem A. and {Bautz}, Mark},
       series = {Proc. SPIE},
       volume = {9144},
        month = jul,
          eid = {914420},
        pages = {914420},
          doi = {10.1117/12.2056811},
       adsurl = {https://ui.adsabs.harvard.edu/abs/2014SPIE.9144E..20A}
}

@INPROCEEDINGS{Gendreu2016,
       author = {{Gendreau}, Keith C. and {Arzoumanian}, Zaven and {Adkins}, Phillip W. and {Albert}, Cheryl L. and {Anders}, John F. and {Aylward}, Andrew T. and {Baker}, Charles L. and {Balsamo}, Erin R. and {Bamford}, William A. and {Benegalrao}, Suyog S. and {Berry}, Daniel L. and {Bhalwani}, Shiraz and {Black}, J. Kevin and {Blaurock}, Carl and {Bronke}, Ginger M. and {Brown}, Gary L. and {Budinoff}, Jason G. and {Cantwell}, Jeffrey D. and {Cazeau}, Thoniel and {Chen}, Philip T. and {Clement}, Thomas G. and {Colangelo}, Andrew T. and {Coleman}, Jerry S. and {Coopersmith}, Jonathan D. and {Dehaven}, William E. and {Doty}, John P. and {Egan}, Mark D. and {Enoto}, Teruaki and {Fan}, Terry W. and {Ferro}, Deneen M. and {Foster}, Richard and {Galassi}, Nicholas M. and {Gallo}, Luis D. and {Green}, Chris M. and {Grosh}, Dave and {Ha}, Kong Q. and {Hasouneh}, Monther A. and {Heefner}, Kristofer B. and {Hestnes}, Phyllis and {Hoge}, Lisa J. and {Jacobs}, Tawanda M. and {J{\o}rgensen}, John L. and {Kaiser}, Michael A. and {Kellogg}, James W. and {Kenyon}, Steven J. and {Koenecke}, Richard G. and {Kozon}, Robert P. and {LaMarr}, Beverly and {Lambertson}, Mike D. and {Larson}, Anne M. and {Lentine}, Steven and {Lewis}, Jesse H. and {Lilly}, Michael G. and {Liu}, Kuochia Alice and {Malonis}, Andrew and {Manthripragada}, Sridhar S. and {Markwardt}, Craig B. and {Matonak}, Bryan D. and {Mcginnis}, Isaac E. and {Miller}, Roger L. and {Mitchell}, Alissa L. and {Mitchell}, Jason W. and {Mohammed}, Jelila S. and {Monroe}, Charles A. and {Montt de Garcia}, Kristina M. and {Mul{\'e}}, Peter D. and {Nagao}, Louis T. and {Ngo}, Son N. and {Norris}, Eric D. and {Norwood}, Dwight A. and {Novotka}, Joseph and {Okajima}, Takashi and {Olsen}, Lawrence G. and {Onyeachu}, Chimaobi O. and {Orosco}, Henry Y. and {Peterson}, Jacqualine R. and {Pevear}, Kristina N. and {Pham}, Karen K. and {Pollard}, Sue E. and {Pope}, John S. and {Powers}, Daniel F. and {Powers}, Charles E. and {Price}, Samuel R. and {Prigozhin}, Gregory Y. and {Ramirez}, Julian B. and {Reid}, Winston J. and {Remillard}, Ronald A. and {Rogstad}, Eric M. and {Rosecrans}, Glenn P. and {Rowe}, John N. and {Sager}, Jennifer A. and {Sanders}, Claude A. and {Savadkin}, Bruce and {Saylor}, Maxine R. and {Schaeffer}, Alexander F. and {Schweiss}, Nancy S. and {Semper}, Sean R. and {Serlemitsos}, Peter J. and {Shackelford}, Larry V. and {Soong}, Yang and {Struebel}, Jonathan and {Vezie}, Michael L. and {Villasenor}, Joel S. and {Winternitz}, Luke B. and {Wofford}, George I. and {Wright}, Michael R. and {Yang}, Mike Y. and {Yu}, Wayne H.},
        title = "{The Neutron star Interior Composition Explorer (NICER): design and development}",
    booktitle = {Space Telescopes and Instrumentation 2016: Ultraviolet to Gamma Ray},
         year = 2016,
       editor = {{den Herder}, Jan-Willem A. and {Takahashi}, Tadayuki and {Bautz}, Marshall},
       series = {Proc. SPIE},
       volume = {9905},
        month = jul,
          eid = {99051H},
        pages = {99051H},
          doi = {10.1117/12.2231304},
       adsurl = {https://ui.adsabs.harvard.edu/abs/2016SPIE.9905E..1HG}
}

@ARTICLE{2014MNRAS.438.2784C,
       author = {{Church}, M.~J. and {Gibiec}, A. and {Ba{\l}uci{\'n}ska-Church}, M.},
        title = "{The nature of the island and banana states in atoll sources and a unified model for low-mass X-ray binaries}",
      journal = {Mon. Not. R. Astron. Soc.},
         year = 2014,
        month = mar,
       volume = {438},
       number = {4},
        pages = {2784-2797},
          doi = {10.1093/mnras/stt2364},
archivePrefix = {arXiv},
       eprint = {1312.1823},
 primaryClass = {astro-ph.HE},
       adsurl = {https://ui.adsabs.harvard.edu/abs/2014MNRAS.438.2784C}
}

@ARTICLE{1989A&A...225...79H,
       author = {{Hasinger}, G. and {van der Klis}, M.},
        title = "{Two patterns of correlated X-ray timing and spectral behaviour in low-mass X-ray binaries.}",
      journal = {Astron. Astrophys.},
         year = 1989,
        month = nov,
       volume = {225},
        pages = {79-96},
       adsurl = {https://ui.adsabs.harvard.edu/abs/1989A&A...225...79H}
}

@article{Church2004,
    author = {Church, M. J. and Bałucińska-Church, M.},
    title = {Measurements of accretion disc corona size in LMXB: consequences for Comptonization and LMXB models},
    journal = {Mon. Not. R. Astron. Soc.},
    volume = {348},
    number = {3},
    pages = {955-963},
    year = {2004},
    month = {03},
    issn = {0035-8711},
    doi = {10.1111/j.1365-2966.2004.07162.x},
    url = {https://doi.org/10.1111/j.1365-2966.2004.07162.x}
    
}

@ARTICLE{2006A&A...460..233C,
       author = {{Church}, M.~J. and {Halai}, G.~S. and {Ba{\l}uci{\'n}ska-Church}, M.},
        title = "{An explanation of the Z-track sources}",
      journal = {Astron. Astrophys.},
         year = 2006,
        month = dec,
       volume = {460},
       number = {1},
        pages = {233-244},
          doi = {10.1051/0004-6361:20065035},
archivePrefix = {arXiv},
       eprint = {astro-ph/0609821},
 primaryClass = {astro-ph},
       adsurl = {https://ui.adsabs.harvard.edu/abs/2006A&A...460..233C}
}

@ARTICLE{2004astro.ph.10551V,
       author = {van der Klis, M.},
  title = {A review of rapid X-ray variability in X-ray binaries},
  journal = {arXiv},
  year = {2004},
  eprint = {astro-ph/0410551}
}

@ARTICLE{1996MNRAS.283..193Z,
       author = {{Zdziarski}, A.~A. and {Johnson}, W.~N. and {Magdziarz}, P.},
        title = "{Broad-band {\ensuremath{\gamma}}-ray and X-ray spectra of NGC 4151 and their implications for physical processes and geometry.}",
      journal = {Mon. Not. R. Astron. Soc.},
         year = 1996,
        month = nov,
       volume = {283},
       number = {1},
        pages = {193-206},
          doi = {10.1093/mnras/283.1.193},
archivePrefix = {arXiv},
       eprint = {astro-ph/9607015},
 primaryClass = {astro-ph},
       adsurl = {https://ui.adsabs.harvard.edu/abs/1996MNRAS.283..193Z}
}

@ARTICLE{1999MNRAS.309..561Z,
       author = {{{\.Z}ycki}, Piotr T. and {Done}, Chris and {Smith}, David A.},
        title = "{The 1989 May outburst of the soft X-ray transient GS 2023+338 (V404 Cyg)}",
      journal = {Mon. Not. R. Astron. Soc.},
         year = 1999,
        month = nov,
       volume = {309},
       number = {3},
        pages = {561-575},
          doi = {10.1046/j.1365-8711.1999.02885.x},
archivePrefix = {arXiv},
       eprint = {astro-ph/9904304},
 primaryClass = {astro-ph},
       adsurl = {https://ui.adsabs.harvard.edu/abs/1999MNRAS.309..561Z}
}

@ARTICLE{Done_2007,
       author = {{Done}, Chris and {Gierli{\'n}ski}, Marek and {Kubota}, Aya},
        title = "{Modelling the behaviour of accretion flows in X-ray binaries. Everything you always wanted to know about accretion but were afraid to ask}",
      journal = {Astron. Astrophys. Rev.},
         year = 2007,
        month = dec,
       volume = {15},
       number = {1},
        pages = {1-66},
          doi = {10.1007/s00159-007-0006-1},
archivePrefix = {arXiv},
       eprint = {0708.0148},
 primaryClass = {astro-ph},
       adsurl = {https://ui.adsabs.harvard.edu/abs/2007A&ARv..15....1D}
}

@ARTICLE{1980A&A....86..121S,
       author = {{Sunyaev}, R.~A. and {Titarchuk}, L.~G.},
        title = "{Comptonization of X-Rays in Plasma Clouds - Typical Radiation Spectra}",
      journal = {Astron. Astrophys.},
         year = 1980,
        month = jun,
       volume = {86},
	pages = {121-138},
       adsurl = {https://ui.adsabs.harvard.edu/abs/1980A&A....86..121S}
}

@article{2017ApJ...850..155S,
       author = {{Shidatsu}, Megumi and {Tachibana}, Yutaro and {Yoshii}, Taketoshi and {Negoro}, Hitoshi and {Kawamuro}, Taiki and {Iwakiri}, Wataru and {Nakahira}, Satoshi and {Makishima}, Kazuo and {Ueda}, Yoshihiro and {Kawai}, Nobuyuki and {Serino}, Motoko and {Kennea}, Jamie},
        title = "{Discovery of the New X-Ray Transient MAXI J1807+132: A Candidate of a Neutron Star Low-mass X-Ray Binary}",
      journal = {Astrophys. J.},
         year = 2017,
        month = dec,
       volume = {850},
       number = {2},
          eid = {155},
        pages = {155},
          doi = {10.3847/1538-4357/aa93f0},
archivePrefix = {arXiv},
       eprint = {1710.03371},
 primaryClass = {astro-ph.HE},
       adsurl = {https://ui.adsabs.harvard.edu/abs/2017ApJ...850..155S}
}

@article{mitsuda1984,
  author = {Mitsuda, K. and Inoue, H. and Koyama, K. and Makishima, K. and Matsuoka, M. and Ogawara, Y. and Suzuki, K. and Tanaka, Y. and Hirano, T.},
  title = {Energy spectra of low-mass binary X-ray sources observed from TENMA},
  journal = {Publ. Astron. Soc. Japan},
  volume = {36},
  pages = {741--759},
  year = {1984}
}

@article{mitsuda1989,
  author = {Mitsuda, K. and Inoue, H. and Nakamura, N. and Tanaka, Y.},
  title = {Spectral variation of luminous X-ray sources in globular clusters},
  journal = {Publ. Astron. Soc. Japan},
  volume = {41},
  pages = {97--118},
  year = {1989}
}

@ARTICLE{white1988,
       author = {{White}, N.~E. and {Stella}, L. and {Parmar}, A.~N.},
        title = "{The X-Ray Spectral Properties of Accretion Disks in X-Ray Binaries}",
      journal = {Astrophys. J.},
         year = 1988,
        month = jan,
       volume = {324},
        pages = {363},
          doi = {10.1086/165901},
       adsurl = {https://ui.adsabs.harvard.edu/abs/1988ApJ...324..363W}
}

@book{Frank2002,
  author    = {Frank, J. and King, A. and Raine, D.},
  title     = {Accretion Power in Astrophysics},
  edition   = {3rd},
  year      = {2002},
  publisher = {Cambridge University Press},
  isbn      = {9780521620536}
}

@article{vanHaaften2012,
	author = {{van Haaften, L. M.} and {Nelemans, G.} and {Voss, R.} and {Wood, M. A.} and {Kuijpers, J.}},
	title = {The evolution of ultracompact X-ray binaries},
	DOI= "10.1051/0004-6361/201117880",
	url= "https://doi.org/10.1051/0004-6361/201117880",
	journal = {Astron. Astrophys.},
	year = 2012,
	volume = 537,
	pages = "A104",
	month = "",
}

@article{Gierlinski2002,
    author = {Gierli{\'n}ski, Marek and Done, Chris},
    title = {A comment on the colour-colour diagrams of low-mass X-ray binaries},
    journal = {Mon. Not. R. Astron. Soc.},
    volume = {331},
    number = {4},
    pages = {L47-L50},
    year = {2002},
    month = {04},
    issn = {0035-8711},
    doi = {10.1046/j.1365-8711.2002.05430.x},
    url = {https://doi.org/10.1046/j.1365-8711.2002.05430.x}

}

@article{Kohlemainen2014,
    author = {Kolehmainen, Mari and Done, Chris and Díaz Trigo, María},
    title = {The soft component and the iron line as signatures of the disc inner radius in Galactic black hole binaries},
    journal = {Mon. Not. R. Astron. Soc.},
    volume = {437},
    number = {1},
    pages = {316-326},
    year = {2013},
    month = {10},
    issn = {0035-8711},
    doi = {10.1093/mnras/stt1886},
    url = {https://doi.org/10.1093/mnras/stt1886}
}

@article{Madsen2017AJ,
  author = {Madsen, K. K. and Harrison, F. A. and Markwardt, C. B. and et al.},
  title = {Calibration of the Nuclear Spectroscopic Telescope Array},
  journal = {Astron. J.},
  volume = {153},
  number = {1},
  pages = {2},
  year = {2017},
  doi = {10.3847/1538-3881/153/1/2}
}

@ARTICLE{Putha2024MNRAS.532.3961P,
       author = {{Putha}, Karthik Gananath and {Bhargava}, Yash and {Bhattacharyya}, Sudip},
        title = "{Probing outbursts of the transient neutron star low-mass X-ray binary Aql X-1 with NICER: a study of spectral evolution}",
      journal = {Mon. Not. R. Astron. Soc.},
         year = 2024,
        month = aug,
       volume = {532},
       number = {4},
        pages = {3961-3971},
          doi = {10.1093/mnras/stae1711},
archivePrefix = {arXiv},
       eprint = {2407.08163},
 primaryClass = {astro-ph.HE},
       adsurl = {https://ui.adsabs.harvard.edu/abs/2024MNRAS.532.3961P}
}

@ARTICLE{Ludlam2018ApJ,
  author = {{Ludlam}, R.~M. and others},
  title = "{NICER Discovers the Ultracompact Orbit of the Accreting Millisecond Pulsar IGR J17062-6143}",
  journal = {Astrophys. J.l},
  volume = {858},
  pages = {L5},
  year = {2018},
  doi = {...}
}

@ARTICLE{Bogdanov2019ApJL,
  author = {{Bogdanov}, S. and others},
  title = "{Constraining the Neutron Star Mass–Radius Relation and Dense Matter Equation of State with NICER. II. Emission from Hot Spots on a Rapidly Rotating Neutron Star}",
  journal = {Astrophys. J.l},
  volume = {887},
  pages = {L26},
  year = {2019},
  doi = {...}
}

@article{remillard2006x,
  title={X-ray properties of black-hole binaries},
  author={Remillard, Ronald A and McClintock, Jeffrey E},
  journal={Annu. Rev. Astron. Astrophys.},
  volume={44},
  number={1},
  pages={49--92},
  year={2006},
  publisher={Annual Reviews},
  doi = {10.1146/annurev.astro.44.051905.092532}
}

@article{Lin_2007,
author = {Lin, D. and Remillard, R. A. and Homan, J.},
  title = {Evaluating Spectral Models and the X-Ray States of Neutron Star X-Ray Transients},
  journal = {Astrophys. J.},
  volume = {667},
  number = {2},
  pages = {1073--1086},
  year = {2007},
  doi = {10.1086/521181}
}

@article{HI4PI2016A&A...594A.116H,
  author = {{HI4PI Collaboration} and {Ben Bekhti}, N. and {Fl{\"o}er}, L. and {Keller}, R. and {Kerp}, J. and {Lenz}, D. and {Winkel}, B. and {Bailin}, J. and {Calabretta}, M. R. and {Dedes}, L. and {Ford}, H. A. and {Gibson}, B. K. and {Haud}, U. and {Janowiecki}, S. and {Kalberla}, P. M. W. and {Lockman}, F. J. and {McClure-Griffiths}, N. M. and {Murphy}, T. and {Nakanishi}, H. and {Pingel}, N. M. and {Stanimirovi{\'c}}, S. and {Stein}, Y. and {Stil}, J. M. and {Taylor}, A. R. and {Tian}, W. W. and {Venzmer}, M. S. and {Walsh}, A. J.},
  title = {HI4PI: A full-sky H I survey based on EBHIS and GASS},
  journal = {Astron. Astrophys.},
  volume = {594},
  pages = {A116},
  year = {2016},
  doi = {10.1051/0004-6361/201629178}
}

@article{Suleimanov2012,
	author = {Suleimanov, V. and Poutanen, J. and Werner, K.},
  title = {X-ray bursting neutron star atmosphere models using an exact relativistic kinetic equation for Compton scattering},
  journal = {Astron. Astrophys.},
  volume = {545},
  pages = {A120},
  year = {2012},
  doi = {10.1051/0004-6361/201219480}
}



\end{document}